\documentclass[prd]{revtex4}

\pdfoutput=1

\newcommand{\be}{\begin{equation}}
\newcommand{\ee}{\end{equation}}
\newcommand{\bea}{\begin{eqnarray}}
\newcommand{\eea}{\end{eqnarray}}

\usepackage{graphicx}
\usepackage{hyperref}

\def\hhref#1{\href{http://arxiv.org/abs/#1}{arXiv:#1}} 

\begin{document}

\title[Heat Kernels and Zeta Functions on Fractals]{Heat Kernels and Zeta Functions on Fractals \footnote{Invited contribution to the JPhysA Special Issue in honour of J. S. Dowker's 75th birthday.}}

\author{Gerald V. Dunne}

\affiliation{Department of Physics, University of Connecticut, Storrs CT 06269 USA}
\begin{abstract}
On fractals, spectral functions such as heat kernels and zeta functions exhibit novel features, very different from their behaviour on regular smooth manifolds, and these can have important physical consequences for both classical and quantum physics in systems having fractal properties.

\end{abstract}

\maketitle

\section{Introduction}
Stuart Dowker has made many fundamental contributions to the modern understanding of heat kernels and zeta functions, and their role in theoretical physics, ranging from quantum field theory on curved space such as De Sitter and anti De Sitter space \cite{dowker-critchley}, to vacuum energy and Casimir physics \cite{dowker-cone}, to finite temperature quantum field theory \cite{dowker-kennedy}. A common theme in his work is the great utility of these spectral functions in encoding physical properties of quantum systems confined to interesting curved manifolds, such as spheres, cones and orbifolds, often highlighting surprising and deeply beautiful connections to number theory. Having personally learned so much from his papers, it is a great pleasure to contribute this paper in honour of his 75th birthday, and to point out some novel features that occur when such smooth manifolds are generalized to fractals. The most direct physical application is to classical and quantum transport in fractal or aperiodic media, which exhibit many novel phenomena \cite{havlin,fibonacci,sornette,meurice,raphael,beenakker,dalnegro}. A more ambitious application is to quantum gravity, both for two-dimensional models \cite{2dgravity}, and in four dimensions. Recently, it has been shown that a wide variety of different approaches to quantum gravity, such as asymptotic safety and the renormalization group, causal dynamical triangulation and  Euclidean dynamical triangulation, all show hints of some fractal-like features of quantum gravity at short distance scales \cite{reuter,loll,coumbe,calcagni,visser}. 

Heat kernels and zeta functions are well known in terms of diffusion processes and path integrals. This is very well understood in flat space, but becomes more interesting and non-trivial when particles and fields reside on curved spaces or couple to gauge fields. The key philosophical idea is to relate spectral information of the propagating fields or particles to the geometric properties of the manifold [or fractal]. It appears that the zeta function is a particularly convenient and efficient way to encode this spectral  information, while the heat kernel enters the physical story primarily  because of its direct relation to the quantum path integral, and to the statistical mechanical partition function.  In a very real sense, we use the propagating fields and particles to probe the spatial geometry. An early example of this, to be elaborated further below, is Lorentz's question: why does the Jeans law only depend on the volume of the region in which a gas, at thermal equilibrium, is contained? This provided physical motivation for  Mark Kac's famous  question: ``Can one hear the shape of a drum?''.  Perhaps the most primitive and basic property of a manifold is its dimension. As is familiar from physics, the behaviour of a system, especially a quantum system, is profoundly dependent on its dimension. Fractals, having fractional and anomalous dimensions, are a natural place to look for exotic physics and mathematics.

The mathematical study of diffusion on fractals was motivated by results from the physics community, where the concepts of spectral dimension and walk dimension were developed \cite{orbach,rammal,kadanoff}. In the late 1980s a general mathematical theory was developed for diffusion and harmonic analysis on fractals \cite{barlow-perkins,carpet,barlow,kigami,strichartz-book,lapidus}. More recently, results have been proved concerning existence and uniqueness of Brownian motion and the Laplacian on a wide variety of fractal spaces. There have also been recent advances proving existence and meromorphic continuations of the zeta function on certain fractals, with rigorous results and explicit expressions for the Sierpinski gasket \cite{teplyaev,derfel}. A general picture is emerging in which the zeta function for highly symmetric fractals has complex poles, and these result in log periodic oscillations in the small time asymptotics of the associated heat kernel trace. Log-periodic oscillations have a long history in physics: in the theory of phase transitions \cite{phase}, the renormalization group \cite{rg}, Levy flights and fractals \cite{montroll}, and generally in systems with a discrete scaling property \cite{derrida}. In this paper, dedicated to Stuart Dowker, I first review the familiar properties of zeta functions and heat kernels on regular smooth manifolds, and then I show how some of these change when generalized to fractals.

\section{Zeta function and heat kernel: basic properties on smooth manifolds}

Before coming to spectral functions on fractals, I review the most relevant basic properties of the zeta function and heat kernel on smooth manifolds \cite{molchanov,elizalde,kirsten,vassilevich}. The later work on fractals will refer to the spectral properties of the Laplacian operator, but we can introduce the main spectral functions for a more general self-adjoint operator $H$. The zeta function is defined as
\begin{eqnarray}
\zeta(s)&=&{\rm tr}\frac{1}{H^s}=\sum_\lambda \frac{1}{\lambda^s} \qquad .
\label{zeta1}
\end{eqnarray}
The sum over the spectrum $\{\lambda\}$ converges for the real part of $s$ large enough, but we are particularly interested in the question of whether or not $\zeta(s)$ has a meromorphic continuation throughout some further region of the complex $s$ plane. For example, in Casimir physics, where the relevant operator is a wave operator, with eigenvalues $\omega^2$, we are interested in making physical sense of the formally divergent sum for the zero point energy:  $\sum \frac{1}{2} \hbar \omega \sim \zeta(-1/2)$. In computing the quantum field theoretic effective action, we need the logarithm of the determinant of the operator $H$ [perhaps a Schr\"odinger operator, or a Klein-Gordon operator, or a Dirac operator, \dots], which is formally related to the derivative of the zeta function at $s=0$, $\ln \det H\sim -\zeta^\prime(0)$, as was first defined by Dowker and Critchley \cite{dowker-critchley}. Making  sense of such formal expressions is the physical  challenge, and the zeta function plays a privileged role.

Another important spectral function, which enters because of its relation to the Euclidean path integral and the thermodynamic partition function, is the  heat kernel trace: 
\begin{eqnarray}
K(t)={\rm tr}\,e^{-H\, t}=\sum_\lambda e^{-\lambda \, t} \qquad .
\label{heat1}
\end{eqnarray}
The heat kernel trace and the zeta function are transforms of one another:
\begin{eqnarray}
\zeta(s)&=& \frac{1}{\Gamma(s)}\int_0^\infty \frac{dt}{t} \, t^s\, K(t)  \qquad ,
\label{transforma}\\
K(t)&=&\frac{1}{2\pi i}\int_{c-i\infty}^{c+i\infty} ds\, t^{-s}\, \Gamma(s)\, \zeta(s) \qquad ,
\label{transformb}
\end{eqnarray}
where the contour for the inverse Mellin transform in (\ref{transformb}) is determined by the analytic properties of the zeta function. 

\subsection{A simple illustrative example: the line segment}

A simple example, which still captures much of the essence of the situation on smooth manifolds, is the Laplacian on the interval of length $\pi$. The eigenvalues of the Laplacian are $\lambda_n=n^2$, with $n$ a non-negative integer starting at $n=0$ for Neumann boundary conditions, but starting at $n=1$ for Dirichlet boundary conditions. Then the heat kernel trace is [$-$ for Dirchlet, and $+$ for Neumann]:
\begin{eqnarray}
K(t)&=&\frac{1}{2}\sum_{n=-\infty}^\infty e^{-n^2\, t}\mp \frac{1}{2}
\nonumber\\
&=&\frac{1}{2}\sqrt{\frac{\pi}{t}}\sum_{n=-\infty}^\infty e^{-n^2\pi^2/ t}\mp \frac{1}{2}
\nonumber\\
&\sim&\frac{\pi}{\sqrt{4\pi t}}\mp  \frac{1}{2}
\label{weyl3}
\end{eqnarray}
The factor $\pi$ in the numerator of the first term is the volume of the one-dimensional interval [cf (\ref{weyl1}) below]. The second term is a boundary contribution, whose sign depends on the type of boundary condition. 
The zeta function is also very simple. For Dirichlet boundary conditions (or, separating the zero mode, a subject to which Stuart Dowker has made  important contributions \cite{dowker-zero}), we see that the zeta function is expressed in terms of the well-known Riemann zeta function:
\begin{eqnarray}
\zeta(s)=\sum_{n=1}^\infty \frac{1}{(n^2)^s}=\zeta_R(2s)
\label{zeta2}
\end{eqnarray}
This zeta function has a well-known analytic continuation into the complex $s$ plane, being meromorphic with a single simple pole at $s=1/2$. From the transform expression (\ref{transformb}) we see that the contour must be taken to the right of the line $Re(s)=1/2$, and when deformed around the poles of $\zeta(s)$ and $\Gamma(s)$ we pick up the two terms of the small $t$ expansion of $K(t)$ in (\ref{weyl3}) [note that the simple poles of $\Gamma(s)$ at the negative integers do not contribute because $\zeta_R(-2n)=0$]. Thus, as can be understood from the transforms (\ref{transforma}, \ref{transformb}), we learn an important lesson:

\begin{equation}
{\rm poles\,\,of\,\,} \zeta(s)\quad  \longleftrightarrow\quad  {\rm small\,\,}\,\, t\,\, {\rm asymptotics\,\,of\,\,} K(t)
\label{poles1}
\end{equation}

\subsection{Smooth Riemannian manifolds: the Weyl expansion}

It is straightforward to generalize the previous example to a $d$-dimensional hypercube, and from there to any smooth Riemannian manifold. This leads to the fundamental result on smooth manifolds, known as the Weyl expansion \cite{molchanov,weyl-review},  that on a $d$-dimensional manifold the heat kernel trace has asymptotic small $t$ behaviour:
\begin{eqnarray}
K(t)\sim \frac{V}{(4\pi t)^{d/2}} - \alpha {S \over (4 \pi t)^{\frac{d-1}{2}}}+\dots
\label{weyl1}
\end{eqnarray}
where $V$ is the volume, $S$ is the boundary volume, and $\alpha$ is a numerical constant determined by the boundary conditions. The dots refer to higher order terms depending on further geometric properties of the manifold such as curvature.
The leading term of the expansion (\ref{weyl1}) relates spectral properties of the Laplacian to basic geometric properties of the manifold: its dimension $d$, and its volume $V$. We can understand the small $t$ asymptotics in (\ref{weyl1}) from Weyl's result \cite{weyl-proof} concerning the large $n$ growth of the eigenvalues, $\lambda_n\sim n^{2/d}$, for a self-adjoint second-order differential operator on a $d$-dimensional smooth manifold:
\begin{eqnarray}
K(t)
\sim \sum_n e^{-n^{2/d}\, t}\sim \int dn\,  e^{- n^{2/d}\, t}\ \sim \frac{c}{t^{d/2}}
\label{weyl2}
\end{eqnarray}
Correspondingly,
\begin{eqnarray}
\zeta(s)\sim \sum_n \frac{1}{(n^{2/d})^s}\sim \zeta_R\left(\frac{2s}{d}\right)
\label{smoothzeta}
\end{eqnarray}
which has a simple pole at $s=d/2$, which, using the transforms in (\ref{transforma}, \ref{transformb}),  is directly related to the leading small $t$ asymptotics in (\ref{weyl2}).

Two natural questions to ask are: on a fractal, is there such a Weyl expansion, and if so, what plays the role of the dimension $d$, and the volume $V$?  These questions are addressed below in Sections 4 and 5.

\subsection{Heat-kernel propagator and off-diagonal Green's function}

Another physically interesting spectral function is the off-diagonal heat kernel, 
\begin{eqnarray}
K(x, y; t)=\langle y | e^{-H t}| x\rangle=P_t(x, y) \qquad ,
\label{od-heat}
\end{eqnarray}
which is the propagator $P_t(x, y)$ for the associated heat equation 
\begin{eqnarray}
\partial_t u = - H u
\label{heat}
\end{eqnarray}
In probabilistic terms, $P_t(x, y)$ is the probability for Brownian motion or diffusion to propagate from $x$ to $y$ in time $t$. In $d$-dimensional flat Euclidean space, the familiar result is 
\begin{eqnarray}
K(x, y; t)=\frac{1}{(4\pi t)^{d/2}}\, \exp\left[-\frac{|x-y|^2}{4t}\right]  \qquad ,
\label{free-propagator}
\end{eqnarray}
which is clearly consistent with (\ref{weyl1}). For a more general manifold, this generalizes \cite{molchanov,liyau}: on a geodesically complete Riemannian manifold with positive Ricci curvature the small $t$ asymptotic behaviour of this propagator is 
\begin{eqnarray}
K(x, y; t)\sim \frac{1}{V(x, \sqrt{t})}\, \exp\left[-c\, \frac{\rho^2(x, y)}{t}\right] +\dots \qquad ,
\label{liyau}
\end{eqnarray}
where $\rho(x, y)$ is the geodesic distance between $x$ and $y$, $c$ is some positive real constant, $V(x, \sqrt{t})$ is the volume of a ball of radius $\sqrt{t}$ centered at $x$, and we assume that $t$ is small enough that there is a unique geodesic connecting $x$ and $y$.

A closely related quantity is the off-diagonal Green's function of the Laplacian:
\begin{eqnarray}
G(x, y)=\langle y | \frac{1}{H}| x\rangle =\int_0^\infty dt\, K(x, y; t) \qquad .
\label{green1}
\end{eqnarray}
On flat space we have the familiar expression
\begin{eqnarray}
G(x, y)=
\begin{cases}
{-\frac{1}{4}\frac{\Gamma\left(\frac{d}{2}-1\right)}{\pi^{d/2}}\, \frac{1}{|x-y|^{d-2}}\qquad, \qquad d\neq 2 \cr
\frac{1}{2\pi}\ln |x-y| \qquad, \qquad d=2
}
\end{cases}
\label{green2}
\end{eqnarray}
Similarly,  on a smooth Riemannian manifold we have the short distance behaviour
\begin{eqnarray}
G(x, y)\sim 
\begin{cases}
{\frac{1}{(\rho(x, y))^{d-2}}\qquad, \qquad d\neq 2 \cr
\ln \rho(x, y) \qquad, \qquad d=2
}
\end{cases}
\label{green3}
\end{eqnarray}
These results highlight the importance in physics of the critical dimension $d=2$, leading to many interesting physical phenomena in planar systems. For propagation on a fractal, we will see below how this critical dimension generalizes in an interesting way.

\subsection{Other spectral functions: Schr\"odinger kernel trace, Poisson kernel trace}

To study both classical and quantum transport on a manifold we often require information about the Schr\"odinger kernel trace
\begin{eqnarray}
S(t)={\rm tr}\,e^{-iH\, t}=\sum_\lambda e^{-i\,\lambda \, t} \qquad ,
\label{schr1}
\end{eqnarray}
which is clearly related to the heat kernel trace $K(t)$, but has completely different convergence properties, being highly oscillatory.  From the resolvent Green's functions
\begin{eqnarray}
G_\pm (E)\equiv \frac{1}{H-E\mp i \epsilon} \qquad, \qquad \epsilon\to 0^+ \qquad ,
\label{gpm}
\end{eqnarray}
we find the density of states 
\begin{eqnarray}
\sigma(E)={\rm tr}\,\delta(H-E)&=&\frac{1}{\pi}\, {\rm Im}\, G_+(E)
\label{dos1}
\end{eqnarray}
Further, defining the S-matrix as 
\begin{eqnarray}
S=\frac{G_+(E)}{G_-(E)}
\label{smatrix}
\end{eqnarray}
we can express the density of states as
\begin{eqnarray}
\sigma(E)=\frac{1}{2\pi i}\frac{\partial}{\partial E}\, {\rm tr}\,\ln\, S
=\frac{1}{2\pi i}\frac{\partial}{\partial E}\, \ln\, \det\, S
\label{dos2}
\end{eqnarray}
On the other hand, the net transmission probability [for one-dimensional transmission] is given by
\begin{eqnarray}
|T(E)|^2=1/\det(G_+(E)\, G_-(E))
\label{transmission}
\end{eqnarray}
Thus, both the density of states and the transmission probability are expressed in terms of the determinant of the resolvent Green's functions, which have the  "proper-time" form in terms of the Schr\"odinger kernel trace:
\begin{eqnarray}
 \ln \det G_\pm (E)={\rm tr}\, \ln G_\pm (E)= -\int_0^\infty \frac{dt}{t}\, e^{\pm i(E\mp i \epsilon)t}\, {\rm tr}\, e^{\mp i H t}
\label{schr2}
\end{eqnarray}

Finally, it is clear that other spectral functions, such as the Poisson kernel trace
\begin{eqnarray}
{\mathcal P}(t)={\rm tr}\,  e^{-\sqrt{H} t}=\sum_\lambda e^{-\sqrt{\lambda}\, t}
\label{poisson}
\end{eqnarray}
can be expressed as a transform of the heat kernel trace
\begin{eqnarray}
{\mathcal P}(t)=\frac{1}{\sqrt{\pi}}\int_0^\infty \frac{du}{\sqrt{u}}\, e^{-u}\, K\left(\frac{t^2}{4u}\right)
\label{pk}
\end{eqnarray}
Similarly, the Poisson kernel trace ${\mathcal P}(t)$ can be expressed in terms of the zeta function, and so its small $t$ asymptotics is also governed by the pole structure of the zeta function.

\section{Some fractal basics: Hausdorff dimension, spectral dimension and walk dimension}

On fractals, the behaviour of these spectral functions is very different. The zeta function $\zeta(s)$ develops {\it complex} poles in the complex $s$ plane, and these lead to log periodic oscillatory behaviour in the small $t$ asymptotics of the heat kernel trace. In the next section we describe these results using an explicit example of the class of {\it diamond} fractals \cite{sasha-eg,adt1}, but we note that these features are quite generic to a wide class of fractal systems \cite{kl,lapidus}. Before discussing this, we recall various natural dimensions associated with fractals.

\begin{enumerate}

\item {\bf Hausdorff-Besicovitch dimension}: this dimension, $d_h$, usually referred to simply as the Hausdorff [or fractal] dimension, is the most familiar dimension associated with a fractal, and refers to the spatial scaling properties of the fractal:
\begin{eqnarray}
d_h=\lim_{r\to 0} \frac{\ln V(r)}{\ln r} \qquad ,
\label{dh}
\end{eqnarray}
where $V(r)$ is the volume of the fractal at length scale $r$. For a smooth manifold $d_h$ reduces to the usual integer-valued dimension $d$, but for a fractal $d_h$ is generically non-integer. For example, for the Sierpinski gasket shown in Figure \ref{fig1}, the Hausdorff dimension is $d_h=\ln 3/\ln 2\approx 1.585$, because on each iteration we halve the length of each interval but the number of intervals grows by a factor of 3.
\begin{figure}[htb]
\centering{\includegraphics[scale=0.5]{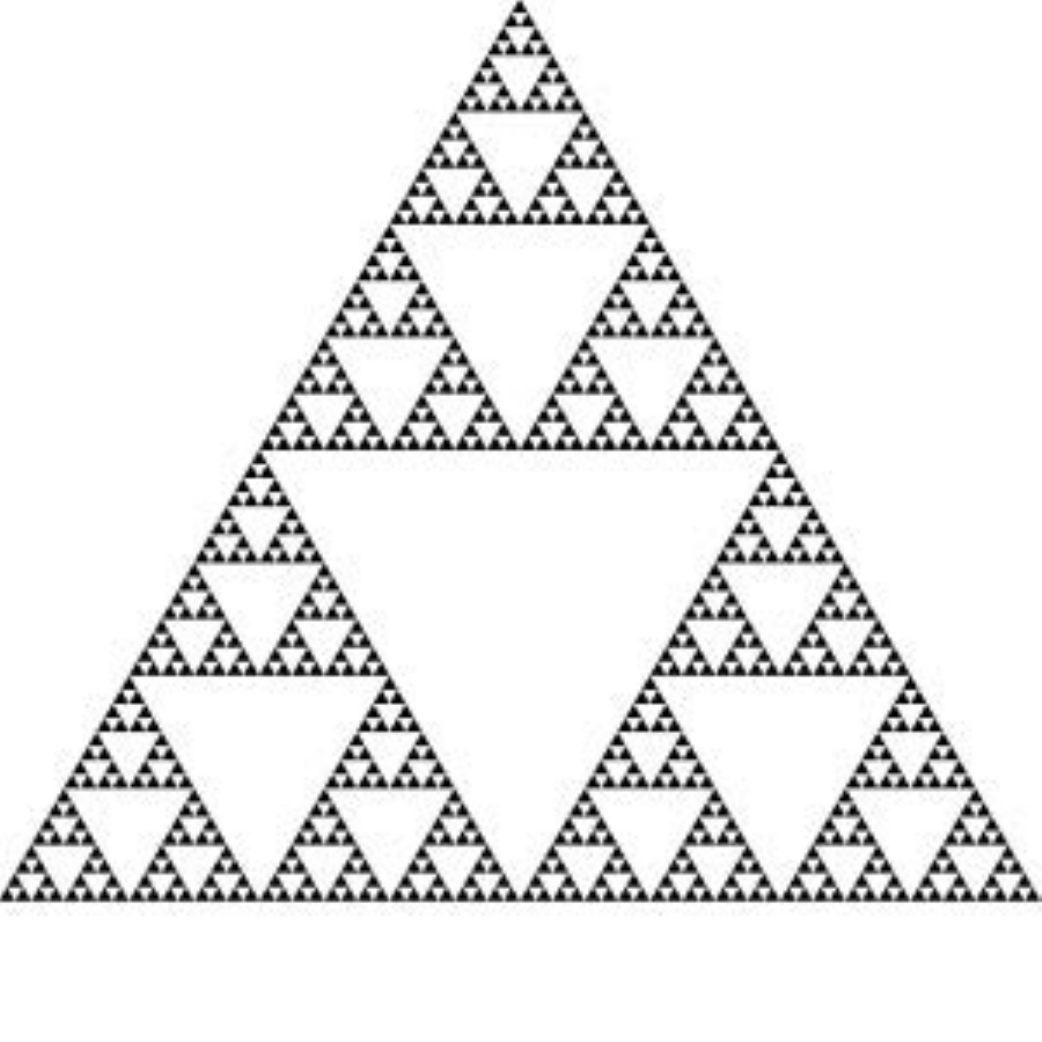}}
\caption{Sierpinksi gasket embedded in two dimensions. This  fractal has Hausdorff dimension $d_h=\ln 3/\ln 2\approx 1.585$, and spectral dimension $d_s=2\, \ln 3/\ln 5\approx 1.365$.}
\label{fig1}
\end{figure}

\item {\bf Spectral dimension}: this dimension, $d_s$, refers to the scaling properties of the eigenvalues of the Laplacian defined on the fractal \cite{orbach,rammal,kadanoff}. It is therefore tied directly to the propagation of heat via the heat equation $\partial_t u=\Delta u$. A simple [but not quite accurate, as we discuss in the next section] way to introduce it is through the small $t$ asymptotics of the heat kernel trace:
\begin{eqnarray}
d_s=-2\lim_{t\to 0} \frac{d\, \ln K(t)}{d\, \ln t} \qquad .
\label{ds1}
\end{eqnarray}
For a smooth manifold, $d_s$ coincides with the integer-valued  dimension $d$, and also with the Hausdorff dimension: $d_s=d_h=d$. However, on a fractal, $d_s$ is generically different from $d_h$, and  also is generically non-integer valued. For example, on the Sierpinski gasket shown in Figure \ref{fig1}, $d_s=2\, \ln 3/\ln 5\approx 1.365$.

\item {\bf Walk dimension}: this dimension, $d_w$, is not really an independent dimension, but is a ratio of dimensions, and has the physical meaning of  a diffusion index \cite{sasha-index}. It encodes the anomalous diffusion that occurs on a fractal, where the usual Einstein relation for Brownian motion is generalized as \cite{orbach,rammal}:
\begin{eqnarray}
\langle r^2(t)\rangle \sim t \qquad \longrightarrow \qquad \langle r^2(t)\rangle \sim t^{2/d_w} \qquad .
\label{dw1}
\end{eqnarray}
Generally, $d_w>2$ on a fractal, so that diffusion spreads more slowly on a  fractal than on a  smooth manifold;  it is called "sub-diffusive" or "sub-Gaussian". In fact, $d_w$ is not independent of the Hausdorff and spectral dimensions. The anomalous diffusion implies that the probability to diffuse a distance $r$ in time $t$ is a scaling function: $p(r, t)=f(r^{d_w}/t)$. But the normalization of total probability involves integrating over the fractal, which necessarily introduces the Hausdorff dimension, written symbolically as: $\int d^{d_h} r\, p(r, t)=1$. This then implies the scaling behaviour
\begin{eqnarray}
p(r, t)=\frac{1}{t^{d_h/d_w}}\, f(r^{d_w}/t) \qquad .
\label{scaling}
\end{eqnarray}
Comparing with the definition of the spectral dimension we see that
\begin{eqnarray}
d_w=\frac{2\, d_h}{d_s} \qquad .
\label{dw2}
\end{eqnarray}
For example, both numerical and analytical work shows that diffusion on the Sierpinski gasket occurs with $d_w=\ln 5/\ln 2\approx 2.32$,  in agreement with the  values of $d_h$ and $d_s$ mentioned above.

\end{enumerate}

\section{Complex poles of the zeta function and log periodic oscillations of  the heat kernel trace}

In fact, for a symmetric fractal, the small $t$ behavior of the heat kernel trace is more involved than the leading term $K(t)\sim1/t^{d/2}$ in (\ref{weyl2}) for a regular smooth manifold. Instead, both the spectral dimension $d_s$ and the Hausdorff dimension $d_h$ play a role in the leading small $t$ behavior.
For very general fractals, the heat kernel trace $K(t)$ is bounded above and below by such a  functional form: there are constants $c_1$ and $c_2$ such that \cite{barlow,kigami}
\begin{eqnarray}
\frac{c_1}{t^{d_s/2}}\leq K(t) \leq \frac{c_2}{t^{d_s/2}} \qquad .
\label{ds2}
\end{eqnarray}
Mathematicians write this in short-hand as
\begin{eqnarray}
K(t) \asymp \frac{c}{t^{d_s/2}} \qquad .
\label{ds3}
\end{eqnarray}
 \begin{figure}[ht]
\centering{\includegraphics[scale=0.35]{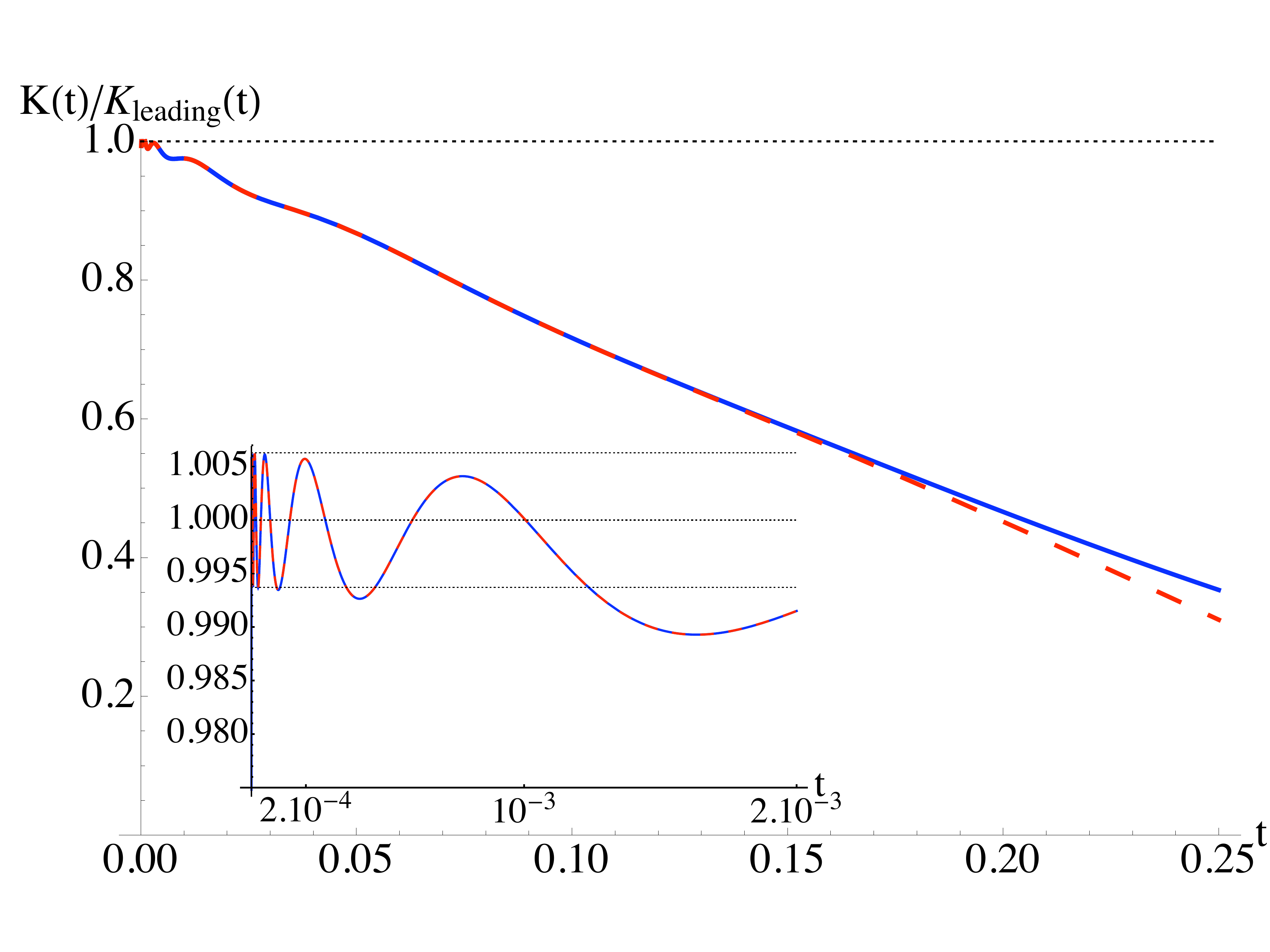}}
\caption{The log periodic oscillations, at small $t$, for the heat kernel trace $K(t)$ on a fractal, normalized relative to the leading term $K_{\rm leading}(t)=c/t^{d_s/2}$ in (\ref{full}, \ref{full2}). The solid [blue] curve is exact; the dashed 
[red] curve is the first two terms in the approximate expression (\ref{full2}).}
\label{fig2}
\end{figure}
For many fractals, those with exact self-similarity [including for example the Sierpinski and diamond fractals],  the result is in fact far more interesting than this, as $K(t)$ exhibits characteristic log periodic oscillations between these upper and lower limits:
\begin{eqnarray}
K(t)\sim \frac{c}{t^{d_s/2}}\left[1+\alpha\, \cos\left(\frac{2\pi}{d_w\, \ln l}\, \ln t+\phi\right)+\dots\right] \qquad ,
\label{full}
\end{eqnarray}
for some real constants $c$, $\alpha$ and $\phi$. Recall that the walk dimension $d_w$ depends on the ratio of the Hausdorff and spectral dimensions, as in (\ref{dw2}). 

\begin{figure}[htb]
\centering{\includegraphics[scale=.4]{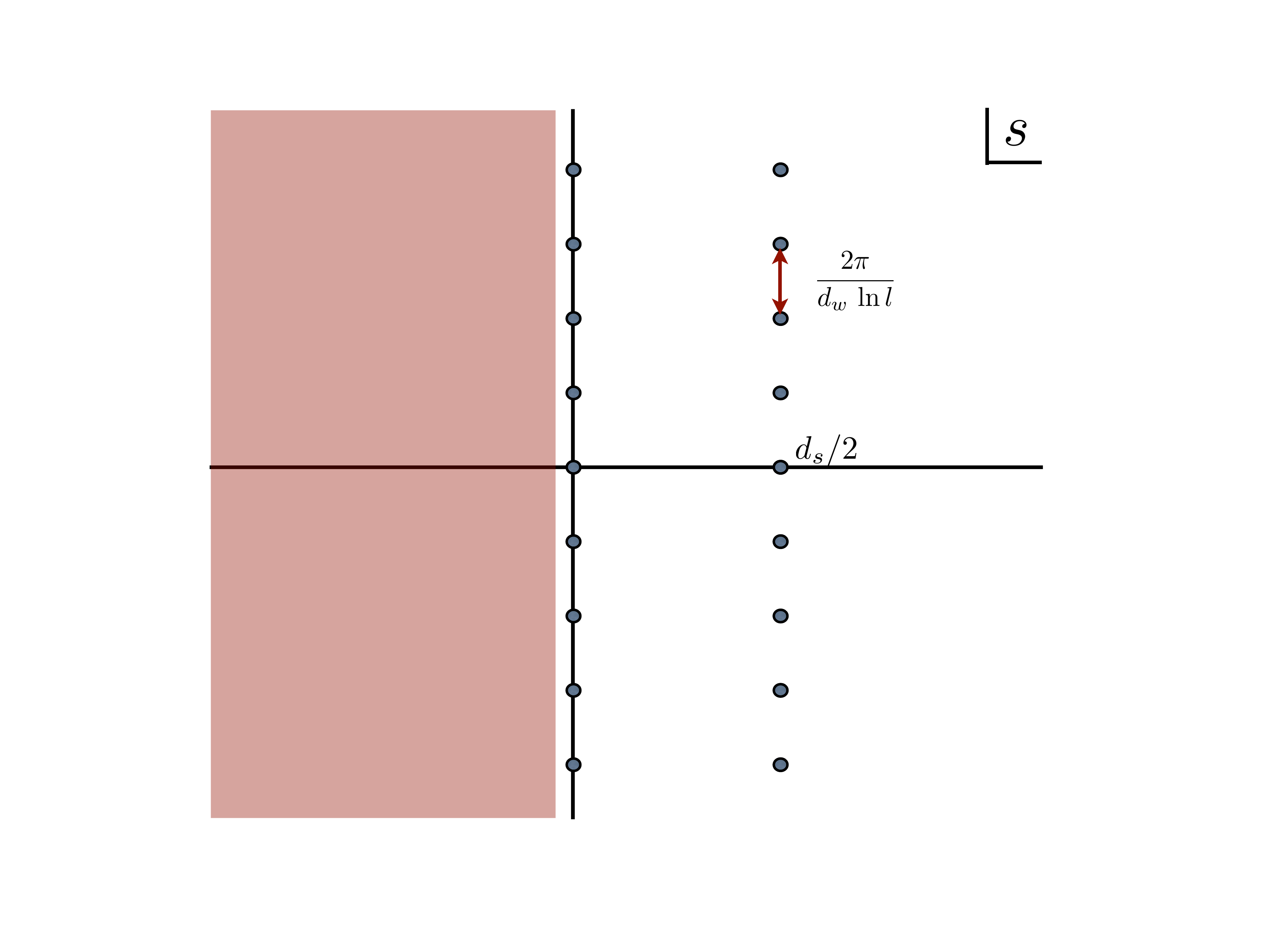}}
\caption{Sketch of the complex pole structure of the zeta function for a fractal. Note the appearance of a tower of complex poles, whose real part defines the spectral dimension $d_s/2$. On certain fractals there may also be a tower of complex poles with ${\rm Re}(s)=0$ \cite{bensasha}, and there is a meromorphic extension of $\zeta(s)$ for all ${\rm Re}(s)>-\epsilon$. Further results on the meromorphic continuation have been obtained by Kajino \cite{kajino}.}
\label{fig3}
\end{figure}

This type of log periodic oscillatory  behavior is show in Figure \ref{fig2}. Note that the log periodic oscillation depends on the walk dimension $d_w$ and on a spatial-decimation factor $l$, which for the Sierpinski gasket is $l=2$ [because in each iteration of the fractal each unit is divided by 2]. Rewriting  (\ref{full}) as
\begin{eqnarray}
K(t)\sim \frac{c}{t^{d_s/2}}+\frac{c\alpha}{2}\left(\frac{e^{-i\phi}}{t^{\frac{d_s}{2}+\frac{2\pi i}{d_w\, \ln l}}}
+\frac{e^{i\phi}}{t^{\frac{d_s}{2}-\frac{2\pi i}{d_w\, \ln l}}}\right)+\dots  \qquad ,
\label{full2}
\end{eqnarray}
and recalling the transforms (\ref{transforma}, \ref{transformb}) that relate $K(t)$ to the zeta function $\zeta(s)$, this form of $K(t)$ can be identified with a tower of {\it complex} poles of the corresponding zeta function:
\begin{eqnarray}
s_m=\frac{d_s}{2}+\frac{2\pi m\, i}{d_w\, \ln l}\qquad, \qquad m\in {\bf Z}
\label{tower}
\end{eqnarray}
as shown in Figure \ref{fig3}.
It turns out that the contribution of the higher complex poles  in the tower is exponentially suppressed by the Gamma function factors coming from (\ref{transformb}), so that (\ref{full2}) gives the leading terms and in fact the second, log periodic, term is much smaller in magnitude than the first term, as can be seen clearly in Figure \ref{fig2}.

Thus,  $d_s$ describes the {\it averaged} leading behaviour of $K(t)$ in the small $t$ limit, while $d_w$ is needed to characterize the log periodic oscillations.
Another definition of the spectral dimension is as the [largest] real part of the complex poles of the zeta function:
\begin{eqnarray}
d_s=2 \max {\rm Re}\left({\rm pole\,\, of}\,\, \zeta(s)\right) \qquad .
\label{ds4}
\end{eqnarray}
We qualify this definition by saying ``max'' because there may be other complex poles, but with smaller real part.
Furthermore, we see that the identification (\ref{poles1}) of the poles of the zeta function with the small $t$ asymptotics of the heat kernel trace generalizes in an interesting way to:
\begin{eqnarray}
\hskip -1 cm{\rm {\bf complex}\,\,poles\,\,of\,\,} \zeta(s)\quad  \longleftrightarrow\quad  {\rm {\bf log\,\,periodic\,\,oscillations}\,\,of\,\,} K(t)
\label{poles2}
\end{eqnarray}

\subsection{Illustrative example: diamond fractals}

\begin{figure}[ht]
\centering{\includegraphics[scale=0.7]{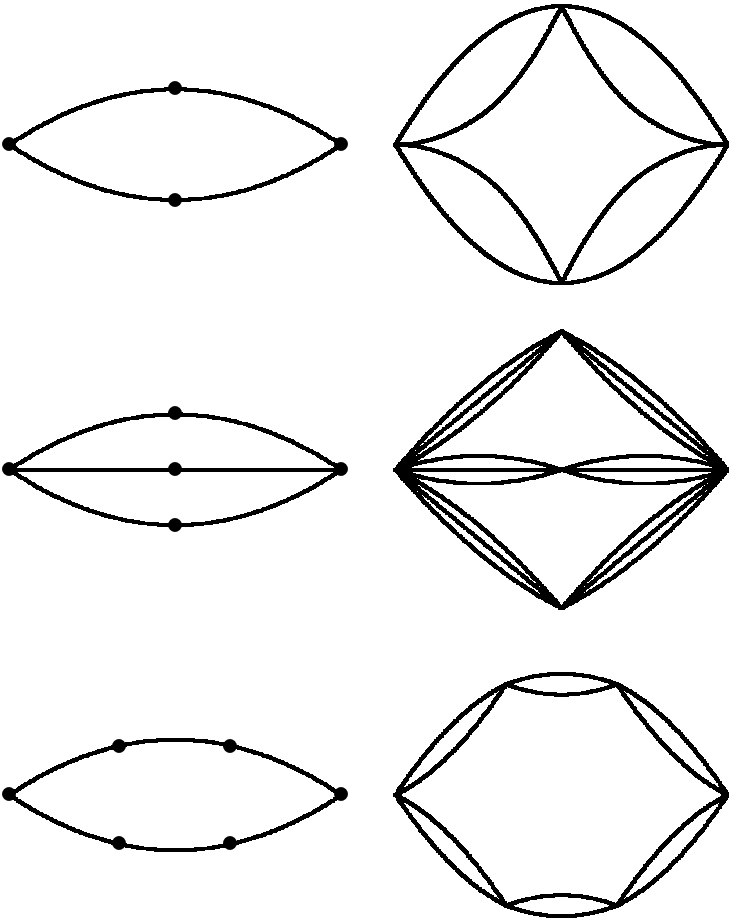}}
\caption{The first 2 iterations of the diamond fractals $D_{4,2}$, $D_{6,2}$ and $D_{6,3}$.  
}
\label{fig4}
\end{figure}

\begin{figure}[htb]
\centering{\includegraphics[scale=1]{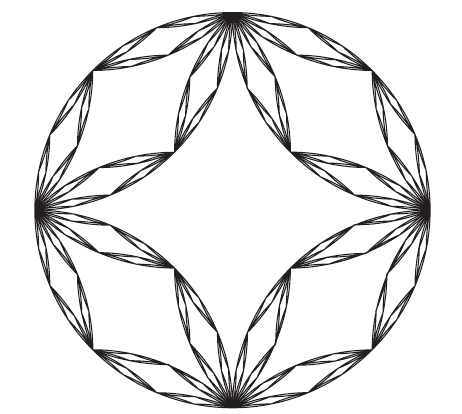}}
\caption{Further iteration of the  diamond fractal $D_{4,2}$.}
\label{fig5}
\end{figure}

These properties of spectral functions on a fractal can be illustrated in explicit detail using the class of fractals known as "diamond fractals", as illustrated in Figures \ref{fig4} and \ref{fig5}. The diamond fractals are defined as iterations of graphs \cite{adt1,sasha-eg}.
At each step $n$ of the iteration, we characterize the diamond fractal by its total length $L_n$, the number of sites $N_n$, and a physical diffusion time $T_n$. Scaling of these  dimensionless quantities allows to define the corresponding Hausdorff $d_h$, spectral $d_s$, and walk $d_w$ dimensions according to
\be
d_h  =  {\ln N_n \over \ln L_n} \quad, \quad {d_s} = 2 {\ln N_n \over \ln T_n} \quad, \quad   {d_w} = {\ln T_n \over \ln L_n} \quad ,
\label{dimensions}
\ee
where the limit $n \rightarrow \infty$ is understood. These three dimensions are thus related by $d_s = 2 d_h/ d_w$ as in (\ref{dw2}).
The diamond fractals are characterized by two integers, one  of which describes how many times each interval is branched, and  the other of which describes the decimation of each interval. Figure \ref{fig4} shows the first two iterations of the diamonds $D_{4,2}$,   $D_{6,2}$ and $D_{6,3}$. The further iteration of  $D_{4,2}$ is shown in Figure \ref{fig5}, which makes more clear the origin of the name ``diamond'' fractal. The associated dimensions are listed in Table \ref{table1}. Notice that by choosing the decimation and branching factors appropriately the spectral dimension can be tuned as desired. Importantly, $d_s$ can be made either smaller than or larger than [or equal to] the critical value 2. Also notice that the diamond fractals are special because they all have walk dimension $d_w=2$, which means that $d_s=d_h$: the spectral and Hausdorff dimensions coincide. This is a special feature, but it does not affect the relevant properties of the spectral zeta function or heat kernel trace.

\begin{table}[hbt]
\begin{center}
\begin{tabular}{||l||r|r|r|c||}
\hline 
& $d_h$ \, \  & $d_w$ \, \ & $d_s = 2 d_h / d_w$ & $l$ \\ 
\hline  
 \, \ \ $D_{4,2}$ & $2$ \, \ \   & $2$ \, \ \
& $2$ \, \ \ & $2$ \, \ \  \\
\hline 
 \, \ \  $D_{6,2}$ & ${\ln 6 / \ln 2} $ & $2$ \, \ \ & ${\ln 6 / \ln 2}$ \, & $2$ \, \ \ \\
\hline
 \, \ \  $D_{6,3}$ & ${\ln 6 /  \ln 3} $ & $2$ \, \ \ & ${\ln 6 / \ln 3} $ \, & $3$ \, \ \ 
\\
\hline 
  $\mbox{Sierpinski}$ & ${\ln 3 / \ln 2}$ & ${\ln 5 / \ln 2}$ &
 $2 {\ln 3 / \ln 5}$ \, &
$2$ \, \ \ \\
\hline 
\end{tabular}
\caption{Fractal dimensions and size scaling factor $l$ for diamond fractals, compared with those  for the Sierpinski gasket. For  $D_{6,2}$, the spectral dimension is $d_s\approx 2.58$, and for $D_{6,3}$, $d_s\approx 1.63$.}
\end{center}
\label{table1}
\end{table}

The heat kernel trace for the Laplacian on a  diamond fractal can be obtained by noticing that the spectrum  is the union of two sets of eigenvalues \cite{sasha-eg,adt1}. One set is composed of the non degenerate eigenvalues $\pi^2 k^2$, (for $k=1, 2, \dots$). This corresponds to the spectrum of the diffusion equation defined on a finite one-dimensional interval of unit length, with Dirichlet boundary conditions. The second ensemble contains iterated eigenvalues, $\pi^2 k^2 L_n ^{d_w}$, obtained by rescaling dimensionless length $L_n$ and time $T_n$ at each iteration $n$ according to $L_n ^{d_w} = T_n$, given in (\ref{dimensions}).  To proceed further, we  use the explicit scaling of the length $L_n = l^n$ upon iteration. These iterated eigenvalues have an exponentially large degeneracy given, at each step, by $B L_n^{d_h} \equiv  B \left( l^{d_h} \right)^n$, where $B =( l ^{d_h -1} -1)$ is known as the  branching factor of the fractal, and the integer $ l^{d_h} $ is the number of links into which a given link is divided.
The {\it exponential} growth of the degeneracy plays a crucial role in our analysis.
Then the diamond heat kernel trace $K_D (t)$ is the sum of  contributions of the two sets of eigenvalues:
\be
K_D (t) = \sum_{k=1}^\infty e^{- k^2 \pi^2 t} + B \sum_{n=0}^\infty l^{n\, d_h} \sum_{k=1}^\infty e^{- k^2 \pi^2 t \, l^{n\, d_w}}  \, . 
\label{hk}
\ee 
Correspondingly, the associated zeta function, $\zeta_D(s)$, is
\bea
\zeta_D (s) &=& {\zeta_R(2s) \over \pi^{2s}} \left( 1 + B \sum_{n=0}^\infty l^{n\,({d_h}  - d_w s)} \right)
 \nonumber \\
&=& {\zeta_R(2s) \over \pi^{2s}} l^{{d_h} -1} \left( {1 - l^{1 - d_w s} \over 1 - l^{{d_h} - d_w s} } \right)\quad ,
\label{zeta3}
\eea
where $\zeta_R(2s)$ is the Riemann zeta function. Note that the zeta function has a factor involving a geometric series;  a very similar structure arises for the zeta function of the Sierpinski gasket \cite{teplyaev,derfel}. The geometric series factor in (\ref{zeta3}) implies that the diamond zeta function $\zeta_D (s)$ has a tower of complex poles given by
\be
s_m = \frac{{d_h}}{d_w} + {2 i \pi m \over d_w \ln l}=\frac{d_s}{2}+ {2 i \pi m \over d_w \ln l} \qquad , \qquad m\in {\bf Z}
\label{poles}
\ee
precisely as in (\ref{tower}). 
These complex poles have been identified with {\it complex dimensions} for fractals \cite{lapidus,teplyaev}, motivated by the form of the small $t$ asymptotics in (\ref{full2}).

Interestingly, the pole of  $\zeta_D(s)$ at $s=1/2$ (coming from the $\zeta_R(2s)$ factor) has zero residue for all diamonds, and so does not contribute to the short time behavior of $K_D(t)$. Remarkably, this vanishing of the residue at $s=1/d_w$  also applies to the analogous zeta function on the Sierpinski gasket \cite{teplyaev}. It is believed that this is generic, but this has not been proved. The pole of $\Gamma(s)$ at $s=0$ gives a constant contribution, $\zeta_D(0)$, to $K_D(t)$.
 \begin{figure}[ht]
\centering{\includegraphics[scale=0.6]{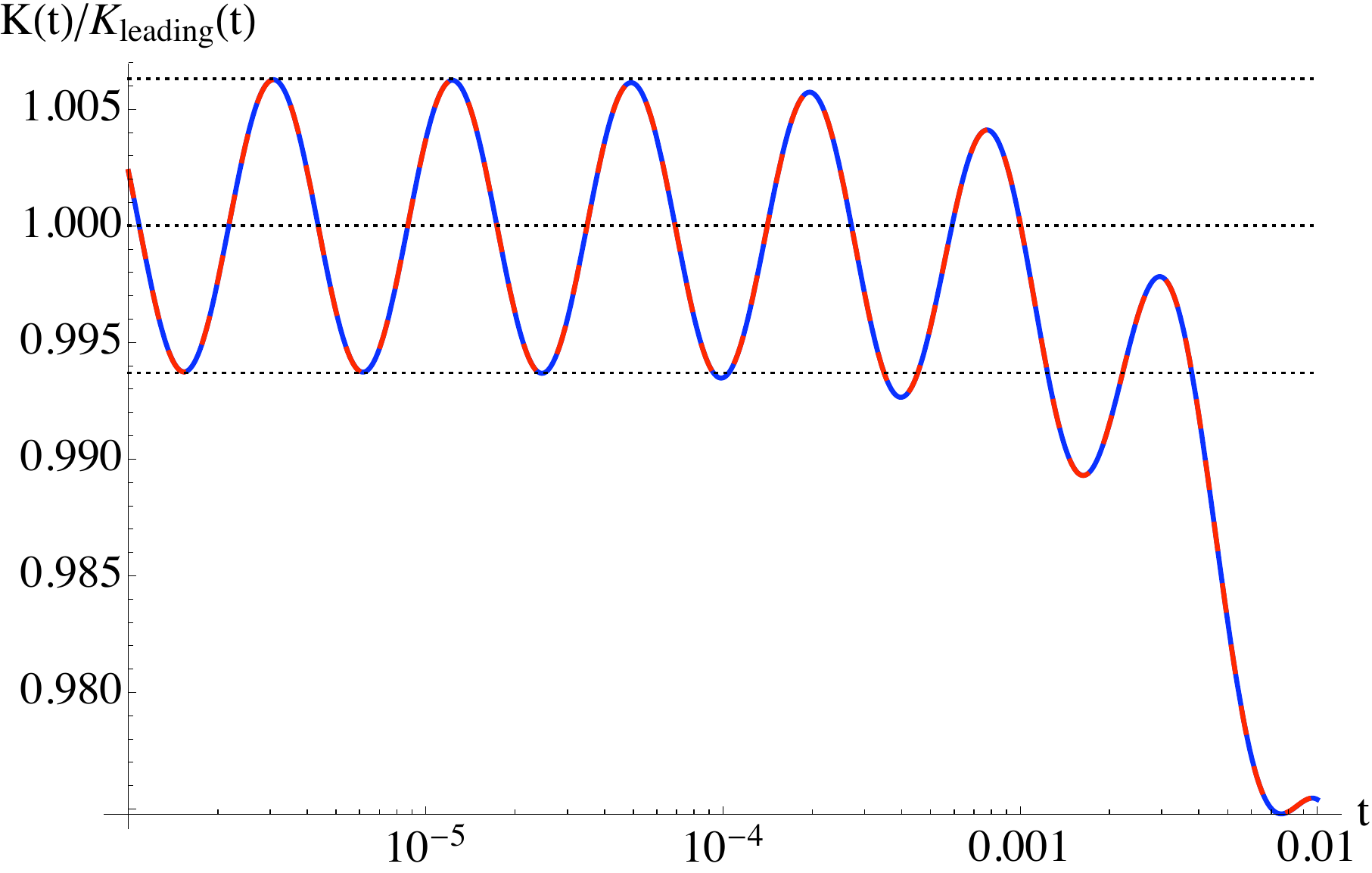}}
\caption{Heat kernel $K_D (t)$ at small time, normalized by the leading non-oscillating term, for the fractal diamond $D_{4,2}$, plotted on a log scale in $t$.  The solid [blue] curve  is exact; the dashed [red] curve is the approximate expression  (\ref{zdetosc}). At such small $t$, these curves are indistinguishable. The relative amplitude of the oscillations  remains constant as $t\to 0$.}
\label{fig6}
\end{figure}
The tower of complex poles in (\ref{poles}), leads to the oscillatory behavior: 
\bea
K_D (t) \sim \left(\frac{l^{d_h -1} -1}{\ln l^{d_w}}\right)\frac{1}{t^{d_s /2}} \left( a_0  + 2 \mbox{Re} \left( a_1 t^{- 2 i \pi/(d_w \ln l) } \right)\right)  +\zeta_D(0)+\dots
\label{zdetosc}
\eea
where the constants are given explicitly by
\begin{eqnarray}
a_m = \frac{\Gamma (s_m) \zeta_R(2s_m)}{\pi^{2s_m}} \qquad .
\label{diamond-coefficients}
\end{eqnarray}
This has precisely the form in (\ref{full}, \ref{full2}), but now in this case of diamond fractals we have explicit expressions for all the constants. 
The leading term $\propto t^{- d_s /2}$ is multiplied by a periodic function of the form $a_{1r} \cos ( \ln t^{s_{1i}} ) + a_{1i} \sin ( \ln t^{s_{1i} })$, where $a_{1r,i}$ are respectively the real and imaginary parts of $a_1$, and $s_{1i} = 2 \pi / \ln l^{d_w}$. These oscillatory terms are small because of the exponentially small amplitudes of the Gamma functions in (\ref{diamond-coefficients}). In Figure \ref{fig6} we compare this leading behaviour with the exact heat kernel trace [dividing out the overall $1/t^{d_s/2}$ factor] and find excellent agreement. The plot is on a log scale in $t$, which makes very clear the log periodic oscillations. This figure also shows that the magnitude of the log periodic oscillations is very small compared to the leading $t$ dependence. Precisely analogous behavior has been found numerically for the Sierpinksi gasket \cite{strichartz}.

\subsection{Intuitive physical explanation of complex poles and log periodic oscillations}

A simple intuitive argument for these novel features of the zeta function $\zeta(s)$ and the heat kernel trace $K(t)$ for fractals is based on the observation  that on fractals the degeneracies of eigenvalues of the Laplacian, and the typical gaps in the spectrum of the Laplacian, both tend to grow {\it exponentially}, rather than {\it polynomially}, as they do on a regular smooth manifold. This is particularly clear on highly symmetric fractals such as the Sierpinski gasket and the diamond fractals, where the eigenfunctions tend to be highly localized and of high degeneracy \cite{sasha-eg}. Contrast this, for example, with the situation on the sphere $S^d$, where the eigenvalues of the Laplacian are 
\begin{eqnarray}
\hskip -1cm \lambda_n=n(n+d-1)\qquad, \qquad {\rm deg}_n=\frac{(2n+d-1)(n+d-2)!}{(d-1)!\, n!}\sim n^{d-1}
\label{sphere}
\end{eqnarray}
and the zeta function has a meromorphic continuation with simple real poles, the largest of which occurs at $s=d/2$ \cite{minak}. Correspondingly, the heat kernel trace has the usual Weyl behaviour in (\ref{weyl1}).

To contrast the consequences of these different large $n$ growths, consider the following ``zeta function'' with general {\it polynomial} growth of degeneracies and eigenvalues, $n^a$ and $n^b$, respectively, with $a$ and $b$ real parameters. [For the Laplacian on a smooth $d$-dimensional manifold we would have $a\sim d-1$, and $b\sim 2$, as for the sphere in (\ref{sphere}).] Then the zeta function is
\begin{eqnarray}
\zeta_{\rm polynomial}(s)=\sum_{n=1}^\infty \frac{n^a}{(n^b)^s}= \zeta_R(b\, s-a) \qquad,
\label{polyzeta}
\end{eqnarray}
which has a simple pole at $s=\frac{(a+1)}{b}$ [which reduces to $d/2$]. The corresponding heat kernel trace function is
\begin{eqnarray}
K_{\rm polynomial}(t)=\sum_{n=1}^\infty n^a\, e^{-n^b\, t} \qquad .
\label{poly}
\end{eqnarray}
A simple Euler-MacLaurin argument yields the small $t$ asymptotics:
\begin{eqnarray}
K_{\rm polynomial}(t)\sim \frac{c}{t^{(a+1)/b}} \qquad ,
\label{poly2}
\end{eqnarray}
in agreement with (\ref{weyl1}) and (\ref{weyl2}). 

\begin{figure}
\centering{\includegraphics[scale=.5]{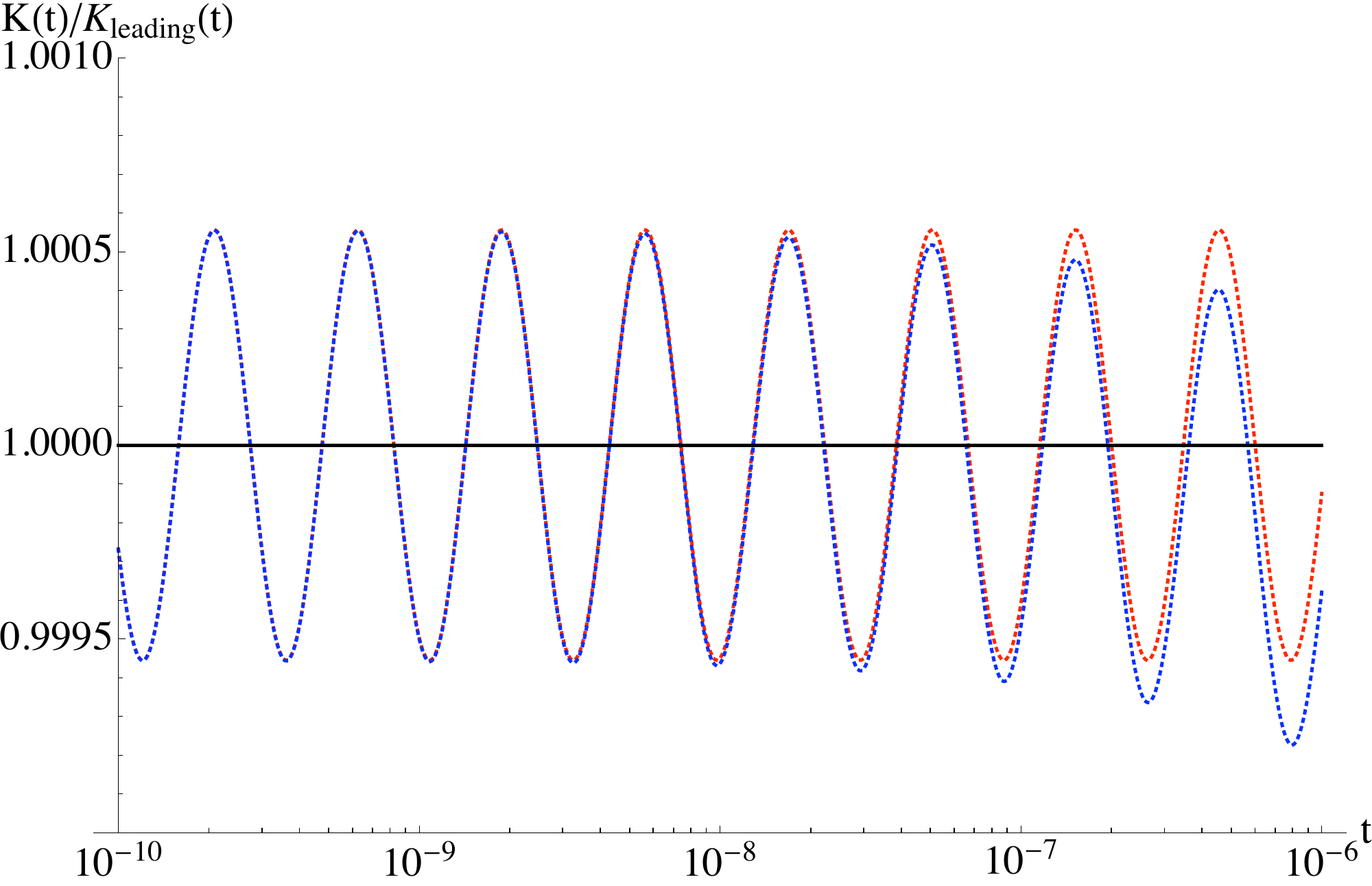}}
\caption{Plot of the ratio of the heat kernel trace function in (\ref{expon}), with exponentially growing degeneracies and eigenvalues,  normalized by its leading small $t$ term. The plots are logarithmic as a function of $t$, and we have chosen parameters $a=2$ and $b=3$. The blue plot shows the function (\ref{expon}) normalized by the leading term in (\ref{weyl-fractal-leading}), while the red plot includes the first log periodic oscillation terms  from (\ref{weyl-fractal-leading}). The agreement is excellent at small  $t$. The log periodic oscillations about the leading behaviour are clearly visible at small $t$, and we also see that these oscillations are of very small amplitude, just as for the diamond fractals shown in Figure \ref{fig6} and for the Sierpinski gasket studied numerically in \cite{strichartz}.}
\label{fig7}
\end{figure}

Now consider the analogous spectral functions with {\it exponentially} growing degeneracies and eigenvalues, $a^n$ and $b^n$, respectively, with $a$ and $b$ real parameters.The zeta function is
\begin{eqnarray}
\zeta_{\rm exponential}(s)=\sum_{n=1}^\infty \frac{a^n}{(b^n)^s}= \frac{\frac{a}{b^s}}{1-\frac{a}{b^s}} \qquad .
\label{exponzeta}
\end{eqnarray}
This zeta function has a tower of complex poles at
\begin{eqnarray}
s_m=\frac{\ln a}{\ln b}+\frac{2\pi i}{\ln b}\qquad, \qquad m\in {\bf Z}
\label{tower-eg}
\end{eqnarray}
just as for the diamond and Sierpinski gasket fractals. Correspondingly, the heat kernel trace function
\begin{eqnarray}
K_{\rm exponential}(t)=\sum_{n=1}^\infty a^n\, \exp\left[-b^n\, t\right]
\label{expon}
\end{eqnarray}
has log periodic oscillatory behaviour:
\begin{eqnarray}
\hskip -2.5cm K_{\rm exponential}(t)\sim \frac{1/\ln b}{t^{\ln a /\ln b}}\left[\Gamma\left[\frac{\ln a}{\ln b}\right]+2{\rm Re}\left( \Gamma\left[\frac{\ln a}{\ln b}+\frac{2\pi i}{\ln b}\right]\exp\left[ -\frac{2\pi i}{\ln b} \, \ln t \right]\right)+ ...\right] 
\label{weyl-fractal-leading}
\end{eqnarray}
In Figure \ref{fig7} we compare this small $t$ asymptotics with the  function $K_{\rm exponential}(t)$, showing clear log periodic oscillations and excellent agreement with this leading form. As for the diamonds and Sierpinski gasket, the contribution from higher complex poles is suppressed by exponentially small Gamma function factors.

This example makes it clear that the physical origin of the log periodic oscillations can be traced to the fact that the degeneracy and eigenvalues grow exponentially with $n$, rather than as a power of $n$, as they do on a smooth Riemannian manifold.

\begin{figure}
\includegraphics[scale=.38]{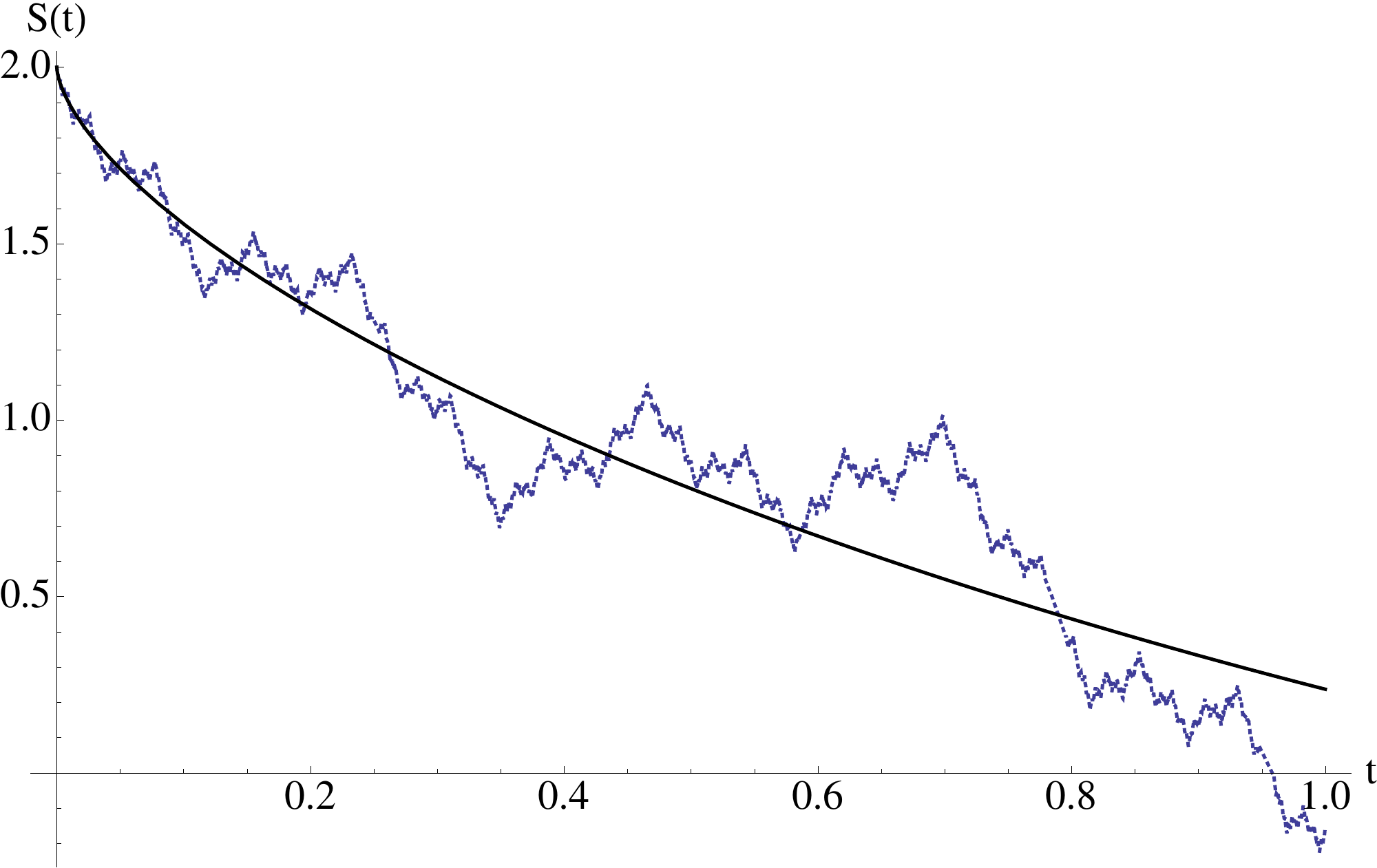}\hskip .2cm
\includegraphics[scale=.38]{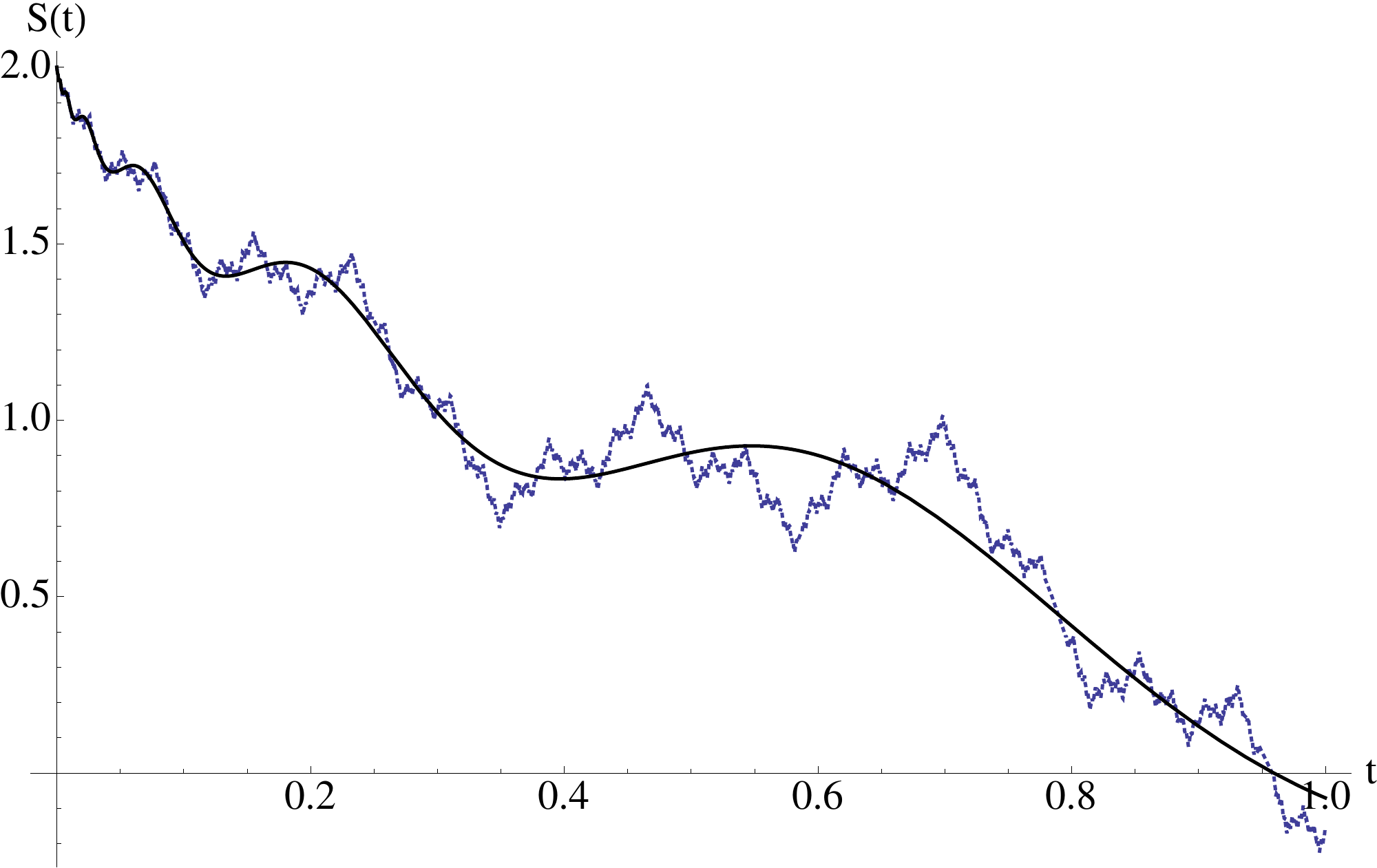}
\includegraphics[scale=.38]{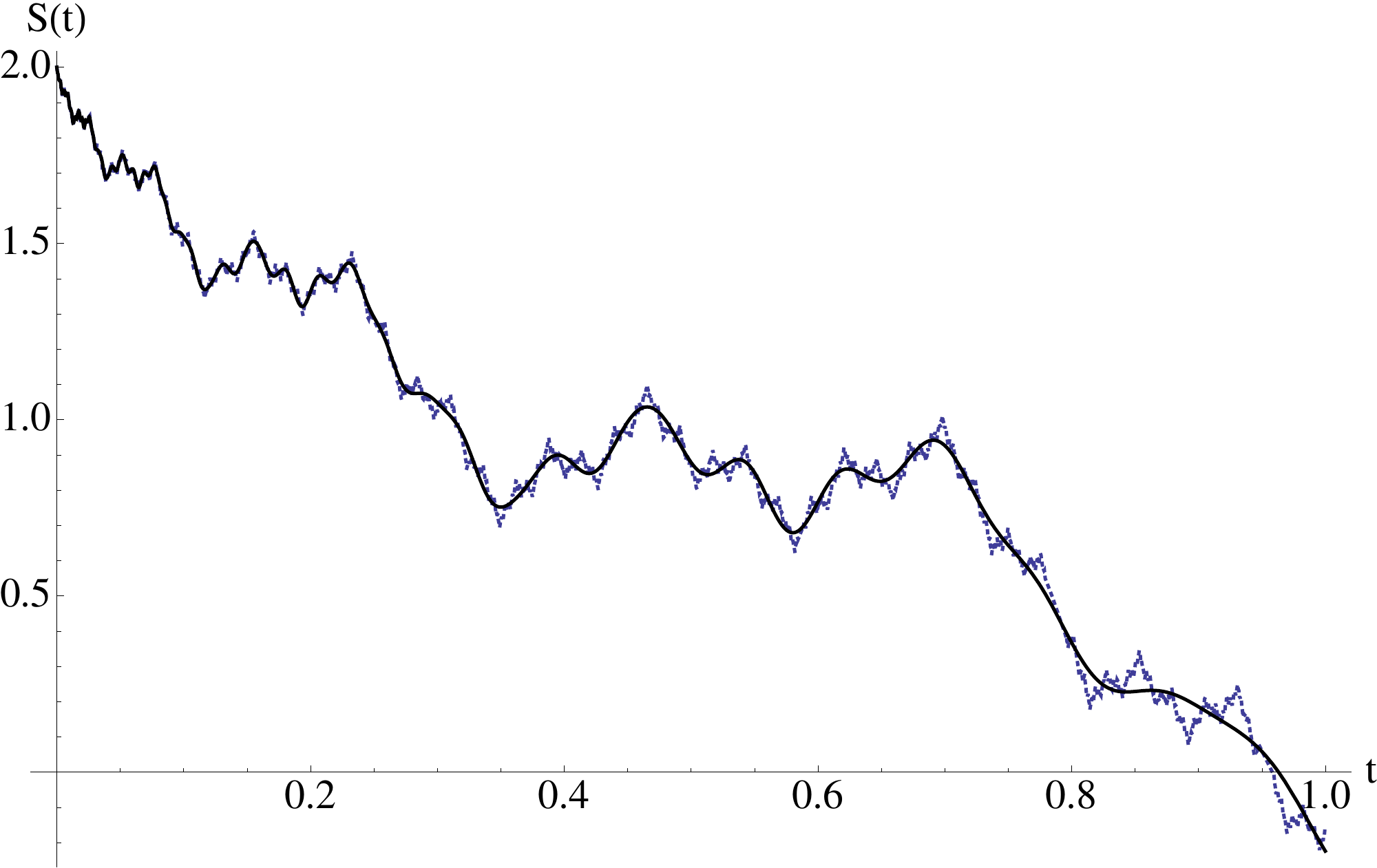}\hskip .2cm
\includegraphics[scale=.38]{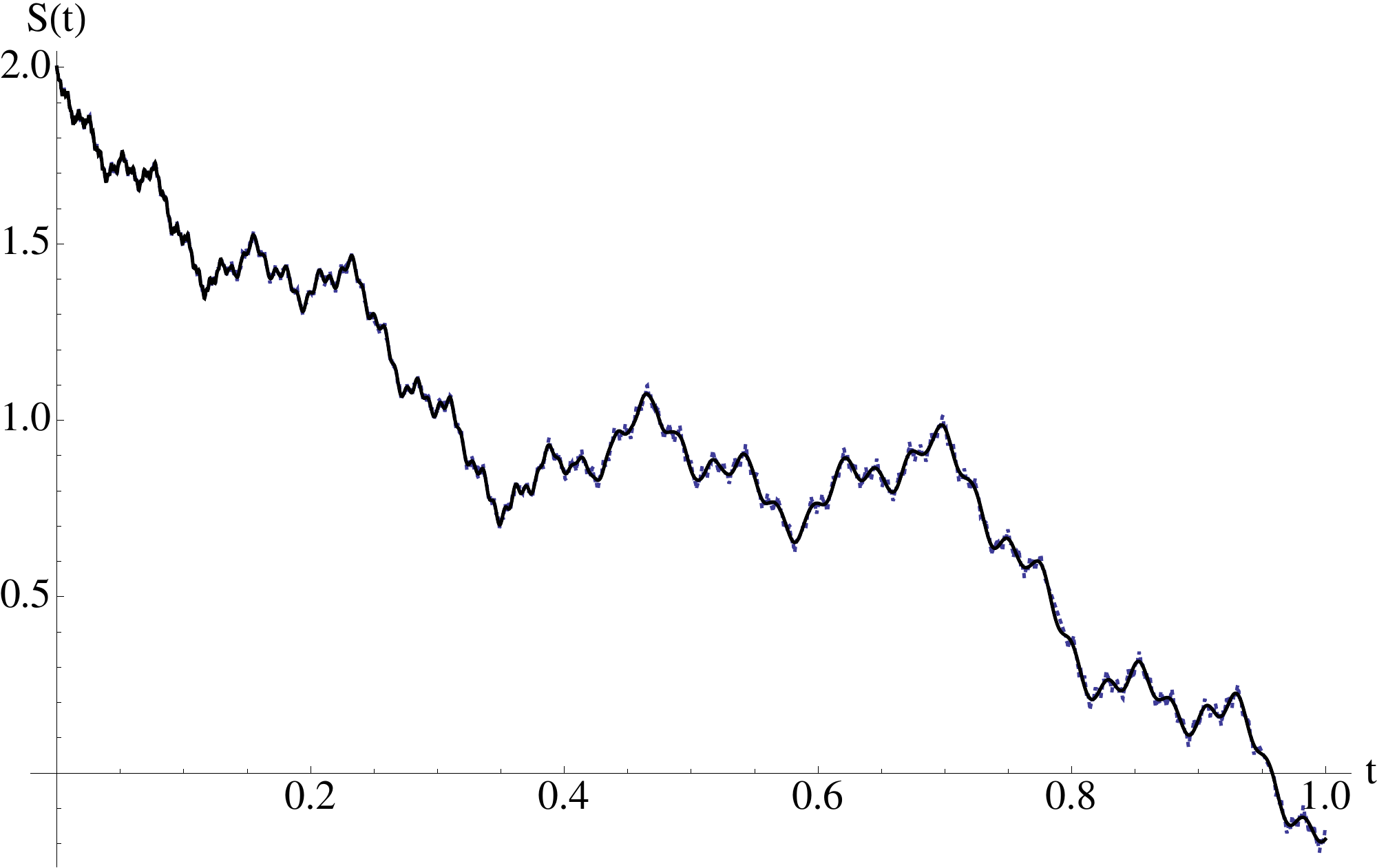}
\caption{Plots of the ``Weierstrass function'' Schr\"odinger kernel $S(t)$ defined in (\ref{expon-schr}), with parameters $a=1/2$ and $b=3$. The blue dotted curve is the exact expression,  while the black solid line in plots (a), (b), (c) and (d) represent the expression in (\ref{expon-schr2}) with the oscillatory $k$ sum up to $|k|$ equal to 0, 1, 10 and 50, respectively. One sees clearly that the agreement is better at smaller values of $t$, and increasing the number of log periodic terms from the $k$ sum improves the agreement out to larger values of $t$. This illustrates the type of wildly oscillatory behaviour expected in a small $t$ expansion of the Schr\"odinger kernel for fractals.}
\label{fig8}
\end{figure}

In the case of the Schr\"odinger kernel (\ref{schr1}) we obtain a highly oscillatory function, whose real part is
\begin{eqnarray}
S_{\rm exponential}(t)=\sum_{n=0}^\infty a^n\, \cos\left(b^n\, t\right)
\label{expon-schr}
\end{eqnarray}
which is precisely the Weierstrass function (we include the $n=0$ term,  by convention). This function appears in the theory of Levy flights and fractals, and the Weierstrass function shows log periodic behaviour \cite{montroll}. This function can be re-written as  \cite{montroll}
\begin{eqnarray}
S_{\rm exponential}(t)&=&  \sum_{n=0}^\infty \frac{(-t^2)^n}{(2n)!\, (1-a\, b^{2n})}\nonumber\\
&&\hskip -4.5cm +\frac{1/\ln b}{t^{\ln a/\ln b}}\sum_{k=-\infty}^\infty \Gamma\left(\frac{\ln a}{\ln b}+\frac{2 \pi i \, k}{\ln b}\right) \cos\left(\frac{\pi}{2}\left(\frac{\ln a}{\ln b}+\frac{2 \pi i \, k}{\ln b}\right) \right) \, \exp\left[-2\pi i\, k\, \frac{\ln t}{\ln b}\right]
\label{expon-schr2}
\end{eqnarray}
which clearly shows the log periodic oscillations. 
This is difficult to plot when $a>1$ because the oscillations are so fast, but we can see in Figure \ref{fig8} the situation for $a=1/2$ and $b=3$. Keeping an increasing number of log periodic terms [corresponding to an increasing number of complex poles of the zeta function], we see better and better agreement with the exact function out to larger and larger  values of $t$. The fractal features are self-evident from the plots.

To conclude this section, we note that this discussion has focused on highly symmetric fractals, those with exact self-similarity, motivated by the fact that these fractals exhibit  novel features such as log periodic oscillations in the heat kernel trace, and associated complex poles of the zeta function. These novel features make them particularly interesting from the physical perspective. But it is important to appreciate that there are other fractals, for example when randomness or disorder is introduced,  where the oscillatory behavior does not necessarily occur. Some specific examples, both mathematical and physical, appear in \cite{kl,hambly}.

\section{Thermodynamics on Fractals}

Motivated by the problem of defining quantum field theory and quantum gravity on a fractal, we consider first a simpler problem: can one define quantum statistical mechanics and thermodynamics on a  fractal space? More precisely, what is the analogue on a fractal of the usual thermodynamic equation of state \cite{lifshitz}
\begin{eqnarray}
P\, V=\frac{U}{d} \qquad ,
\label{eos1}
\end{eqnarray}
for a smooth $d$-dimensional closed manifold of volume $V$, with pressure $P$ and internal energy $U$? It is not immediately clear on a fractal space which dimension plays the role of $d$, or what is the appropriate volume $V$, because naively the volume of a fractal is infinite.  As mentioned already in the introduction, Mark Kac notes in his famous paper  ``Can One Hear the Shape of A Drum?'' \cite{kac}, that an earlier version of this question had been posed by Lorentz for a regular manifold, who asked why the Jeans law of thermodynamics only refers to the total volume of the region occupied by a gas, and not to its shape. Here we pose the analogous question for a fractal spatial region. The physical idea is to use propagating
particles, as they achieve thermal equilibrium, to probe the geometry of the region to which they are confined.

Appropriately for this article, the most efficient way to approach this problem was pioneered by  Dowker and Kennedy \cite{dowker-kennedy}, who studied quantum field theory on  spacetimes of the form $S^1\times \mathcal M$, with $S^1$ being the compactified Euclidean time of period $\hbar \beta=\hbar/T$ for the Matusbara modes. To illustrate this idea consider a massless bosonic field [a ``photon''] confined to live on a spatial region that is a fractal $\mathcal F$. The thermodynamic partition function is expressed as
\begin{equation}
\ln \mathcal Z (T,V) =-\frac{1}{2} \ln \mbox{Det}_{S^1 \times \cal F} \left( \partial ^2 / \partial t^2 - c^2 \Delta \right)
\label{part2}
\end{equation}
where $\Delta$ is  the Laplacian  on the spatial region $\cal F$, and $c$ is a velocity parameter [the ``speed of light'']. 
The expression (\ref{part2}) is usually derived using a mode decomposition of the bosonic field, summing over states in phase space, which in turn relies on an analysis of Fourier transforms. But on a fractal there is, as yet, no well-defined notion of Fourier analysis. However, we can by-pass this step as follows. 
Recall that the spectral partition function for a single bosonic oscillator of frequency $\omega$ is
\bea
\ln Z (T,\omega)  &=&- { \beta \hbar \omega \over 2} - \ln \left( 1 - e^{- \beta \hbar \omega} \right) \nonumber\\
&=& - { \beta \hbar \omega \over 2}  +\sum_{n=1}^\infty \frac{1}{n}e^{-\beta \hbar \omega n} \nonumber\\
&=& - { \beta \hbar \omega \over 2}  +\sum_{n=1}^\infty \frac{\beta \hbar}{\sqrt{4\pi}}\int_0^\infty \frac{d\tau}{\tau^{3/2}} \,e^{-\omega^2 \tau}\, e^{-(n\beta \hbar)^2/(4\tau)} \nonumber\\
&=& \frac{1}{2} \int_0^\infty {d \tau \over \tau} e^{- \omega^2 \tau}  \sum_{n = - \infty}^{+ \infty} e^{-  ({2 \pi n\over \hbar \beta})^2 \tau}
\label{eq3}
\eea
where in the last step we have used the Poisson summation formula. Using the  Kubo, Martin and Schwinger (KMS) condition \cite{kms} for thermodynamic equilibrium of a bosonic field $\phi$ at temperature $T$, we recognize the second factor in the integrand as a sum over Matsubara frequencies $(\frac{2 \pi n}{\hbar \beta})^2$, corresponding to the spectrum of the operator $\partial^2/ \partial t^2$ with periodic boundary conditions $\phi (t + \hbar \beta ) = \phi (t)$.
Identifying $\omega^2=c^2 k^2$ with the eigenvalues of $-c^2\Delta$, and tracing over all modes  \footnote{Note: $\ln \mathcal O=-\int_0^\infty \frac{d\tau}{\tau}e^{-\mathcal O\, \tau}$, and $\ln {\rm Det}\, \mathcal O={\rm Tr}\,\ln  \mathcal O$.}, we recover expression (\ref{part2}), which can be written as follows 
\be
\ln \mathcal Z (T,V)  = \frac{1}{2} \, \int_0^\infty \frac{d \tau}{\tau}  \, \Theta_3\left(0, e^{-\left(\frac{2\pi}{L_\beta}\right)^2\tau}\right)\, K_{\cal F}(\tau) \qquad ,
\label{partition-hk}
\ee
where $\Theta_3$ is the Jacobi theta function and we have defined the "photon" thermal  wavelength, $L_\beta\equiv \beta\hbar c$.
All  information about the spatial region ${\mathcal F}$, including possible dependence on its dimension and  volume, is contained in the heat kernel trace factor $K_{\cal F}(\tau)$, while the thermodynamic information is encoded in the theta function factor. The form in (\ref{partition-hk}) is particularly convenient because it factors, in the integrand, the temperature dependence from the spectral dependence of the Laplacian on the spatial region.

The heat kernel $K_{\cal F}(\tau)$ contains an implicit length scale $L_s$, which we call the ``spectral length'', 
that defines a dimensionless Laplacian $\tilde{\Delta}\equiv L_s^2\, \Delta$ [recall that 
the eigenvalues of $\Delta$ have dimensions of $1/{\rm length}^2$]. Results in the mathematical literature refer to the dimensionless Laplacian $\tilde{\Delta}$. For black-body radiation confined to manifolds of complicated shape, it is difficult to make an explicit mode decomposition and  find an explicit expression for the heat kernel trace, except for highly symmetric regions such as hyper-cubes or spheres and sections of spheres \cite{dowker-sphere}. However, we can learn about the thermodynamic [large volume] limit from the Weyl expansion of the heat kernel trace. Note that the large volume thermodynamic  limit corresponds to $L_s \gg  L_\beta $, which is a high temperature limit $T \gg \hbar c /L_s$. This probes the small $\tau$ behavior of $\mathcal K_{\mathcal F}(\tau/L_s^2)$. Using (\ref{full2}), the leading term is
 \be
 {\cal K}_{\mathcal F} (\tau/L_s^2)  \sim \frac{c\, L_s^{d_s}}{(4 \pi \tau )^{d/2}} + \cdots  \qquad ,
 \label{weyl-thermal}
 \ee
 which we can use to {\it define} the ``spectral volume'' $V_s$ \cite{adt2}
 \begin{eqnarray}
 V_s=c\, L_s^{d_s}  \qquad ,
 \label{vs1}
 \end{eqnarray}
 where the numerical coefficient is determined below [see (\ref{vs2})]. Physically, we are using the photons in thermal equilibrium to {\it define} the volume of the spatial region in which they are confined. This physical idea gives a direct link between the thermodynamic limit and the leading term of the Weyl expansion:
 \begin{eqnarray}
{\rm thermodynamic\,\, limit} \qquad  \leftrightarrow\qquad  {\rm leading\,\,term\,\,of\,\,Weyl\,\,expansion}
\label{thermo-weyl}
\end{eqnarray}
For photons confined to a regular manifold region, this  reproduces all the standard results of thermodynamics, including the equation of state (\ref{eos1}). To see this, recall the thermodynamic limit following from the leading Weyl behaviour (\ref{weyl1}):
\begin{eqnarray}
\ln\mathcal Z &\sim &\frac{\zeta_R(d+1)\Gamma\left(\frac{d+1}{2}\right)}{\pi^{(d+1)/2}} \left(\frac{V}{L_\beta^d}\right)  \nonumber\\
P&=&{1 \over \beta} \left( {\partial \ln \mathcal Z \over \partial V} \right)_T
\sim \frac{\zeta_R(d+1)\Gamma\left(\frac{d+1}{2}\right)}{\pi^{(d+1)/2}} \left(\frac{T}{L_\beta^d}\right)  
\nonumber\\
U&=&- {\partial \over \partial \beta} \ln \mathcal Z (T,V) 
\sim  \frac{\zeta_R(d+1)\Gamma\left(\frac{d+1}{2}\right)}{\pi^{(d+1)/2}} \left(\frac{d}{\hbar \, c\, L_\beta^{d-1}}\right)  
\label{thermo1}
\end{eqnarray}
This leads immediately to the usual equation of state (\ref{eos1}).

We can also formulate this in terms of the zeta function $\zeta(s)$ (see again the work of Dowker and Kennedy \cite{dowker-kennedy}). For a regular manifold ${\mathcal M}$, straightforward manipulations show that the partition function can be expressed as 
\bea
\hskip -2cm \ln \mathcal Z(T, V)=-\frac{1}{2}\left(\frac{L_\beta}{L}\right)\,\zeta_{\mathcal M}\left(-\frac{1}{2}\right)
+\frac{1}{\pi i}\int_C \left(\frac{L}{L_\beta}\right)^{2s}\, \Gamma(2s)\, \zeta_{R}(2s+1)\, \zeta_{\mathcal M}(s) \, ds
\label{zetaform}
\eea
The first term gives the standard zero temperature ``vacuum energy" contribution \cite{dowker-kennedy}, proportional to 
$\zeta_{\mathcal M}\left(-\frac{1}{2}\right)$,
which has recently been generalized to quantum graphs  \cite{harrison}. 
The second term in (\ref{zetaform}) encodes  finite temperature corrections to the internal energy $U$; the Riemann zeta  factor $\zeta_{R}(2s+1)$ arises from the sum over Matsubara modes. The large volume behaviour is controlled by the largest pole of $\zeta_{\mathcal M}(s)$, which occurs at $s=d/2$, and reproduces the thermodynamic limit in (\ref{thermo1}), and also determines the volume $V$ in terms of the zeta function:
\begin{eqnarray}
V=(4\pi)^{d/2}\Gamma(d/2) L^d\, {\rm Res}\left[\zeta_{\mathcal M}(s)\right]_{s=d/2}
\label{volume}
\end{eqnarray}

On a fractal spatial region ${\mathcal F}$ the same procedure follows through, replacing $\zeta_{\mathcal M}(s)$ with $\zeta_{\mathcal F}(s)$.  Together with the definition of the spectral volume in (\ref{vs1}),  the equation of state is found to be  \cite{adt2}
\be
P \, V_s \, = \, \frac{1}{d_s} \, U
\label{eos2}
\ee
This has the same form as (\ref{eos1}), but physically is quite different. 
First, note that the  spectral volume scales with the length $L_s$ to the power $d_s$, the spectral dimension, not the Hausdorff dimension. Consequently, the dimension $d$ in (\ref{eos1}) is replaced by the spectral dimension $d_s$ in the fractal equation of state (\ref{eos2}). The numerical coefficient in the spectral volume (\ref{vs1}) is fixed by the residue of the zeta function at the pole at $s=d_s/2$:
\begin{eqnarray}
V_s=(4\pi)^{d_s/2}\Gamma(d_s/2) L_s^d\, {\rm Res}\left[\zeta_{\mathcal F}(s)\right]_{s=d_s/2}
\label{vs2}
\end{eqnarray}

Given that there is no well-defined notion of Fourier transforms on fractals, it is somewhat surprising that this thermodynamic analysis is possible. A standard approach to thermodynamic equilibrium is in terms of momentum modes, as we write
\begin{eqnarray}
\ln {\cal Z} =-\frac{\beta}{V_{\rm mom}} \int\frac{d^d k}{(2\pi)^d} 
\left(\frac{ \hbar \omega}{2}+\frac{1}{\beta} \ln \left(1-e^{-\beta \hbar \omega}\right)\right)
\label{mom-modes}
\end{eqnarray}
where we have explicitly written $V_{\rm mom}$ for the momentum space volume. On a regular smooth manifold $V_{\rm mom}\sim 1/V$, the inverse of the spatial volume, and so  $V$ enters the thermodynamic equation of state. On the other hand, it is clear that the $V$ in the equation of state is really $1/V_{\rm mom}$, and on a fractal this is replaced by the spectral volume $V_s$ defined in (\ref{vs2}). This suggests a provocative idea, which deserves to be explored further: on a fractal-like region, it appears that the spectral volume (which we might identify with momentum space) has, in some appropriate sense, a dimension of $d_s$, while the spatial volume has a dimension of $d_h$. On a smooth manifold these dimensions are the same, but on a fractal they are generically different. This could lead to some potentially unusual properties of thermodynamic equilibrium, which deserve to be probed experimentally.

\section{Approximate Fractals}

\begin{figure}[htb]
\centering{\includegraphics[scale=0.8]{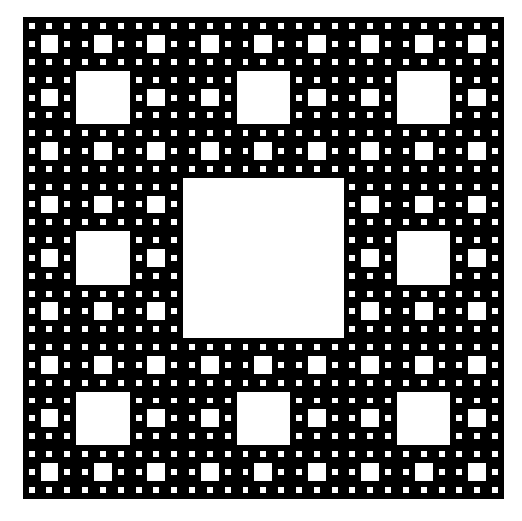}}
\caption{A Sierpinksi carpet. Its Hausdorff  dimension is $d_h=\ln 8/\ln 3\approx 1.89$, but its spectral dimension $d_s$ is only known numerically, estimated as $d_s\approx 1.80525$, and rigorously  bounded as $1.674 < 2\ln8/\ln 12\leq d_s \leq 2\ln8/\ln(28/3) <1.86$ \cite{bass-carpet}.}
\label{fig9}
\end{figure}

The possibility of probing experimentally such strange properties of fractal regions raises an immediate objection that one can never fabricate an infinitely iterated fractal such as a perfect Sierpinski gasket or diamond,  or even a Sierpinski carpet [see Figure \ref{fig9}], because there is always a lower limit on the spatial scale that can be realized in a real physical  system. However, it turns out that these strange properties [for example, the spectral dimension and log periodic oscillations]  can be seen already with relatively modest level of iteration of the fractal. It is a question of resolution scale, and can be illustrated as follows. 

\begin{figure}[htb]
\centering{\includegraphics[scale=0.5]{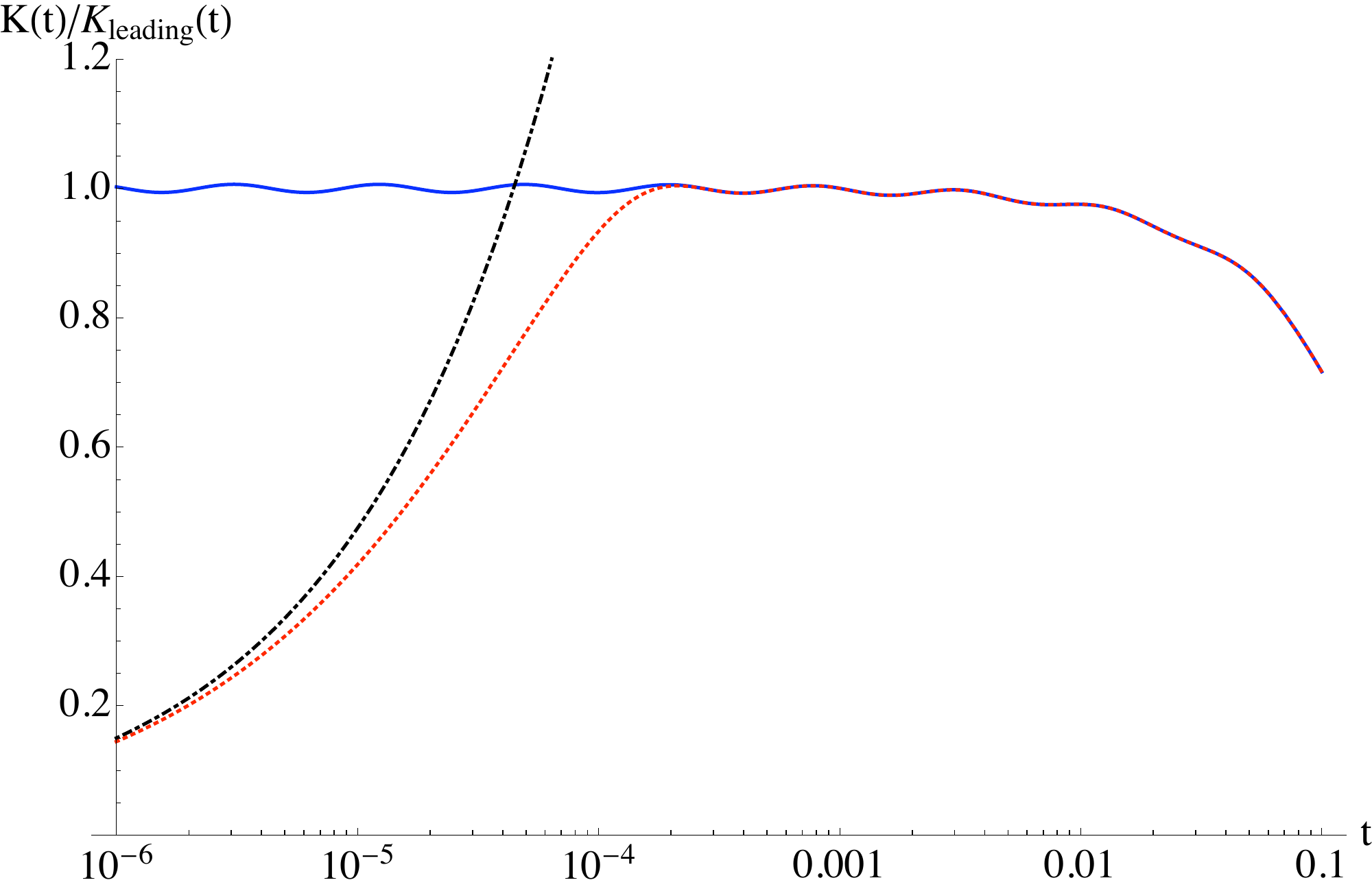}\\
\includegraphics[scale=0.5]{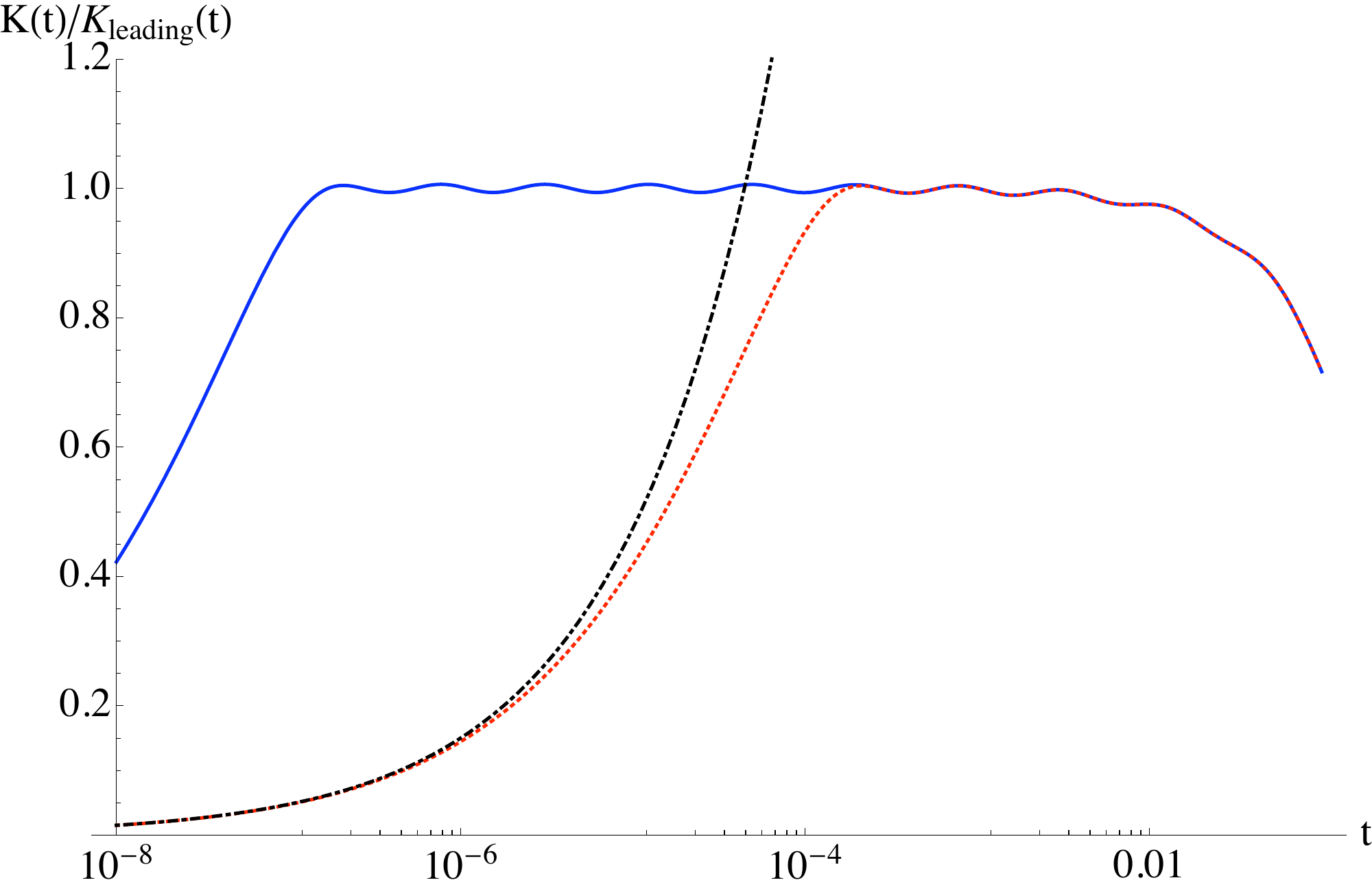}}
\caption{Plots of the heat kernel trace $K(t)$, normalized by the leading fractal form in (\ref{full2}), for the $D_{6,2}$ diamond fractal, plotted on a log scale in  $t$. The solid [blue] curves show the result for 10 iterations of the fractal, while the dotted [red] curves show the result for 5 iterations.  The dot-dashed [black] curves show the expected result for a one-dimensional quantum graph. The first plot probes $t$ down to $10^{-6}$, while the second probes $t$ down to $10^{-8}$. Notice the region of $t$ in which these approximate fractals behave as if they had spectral dimension $d_s=\ln 6/\ln 2\approx 2.58$, and also exhibit log periodic oscillations with the expected period. Also note that this region extends down to smaller values of $t$ as the number of iterations increases.}
\label{fig10}
\end{figure}

To be very explicit, we consider again the diamond fractals. For a finite level of iteration, a diamond fractal is undoubtedly a one-dimensional quantum graph. Therefore, in the asymptotic small $t$ limit, the heat kernel trace should behave in a way governed by the one-dimensional Weyl law (\ref{weyl1}), rather than by the fractal Weyl law (\ref{full2}). This is shown in Figure \ref{fig10}, where we plot the ratio of the heat kernel trace $K(t)$ to the leading fractal value from (\ref{full}), for 5 and 10 levels of iteration of the diamond. This is plotted as a function of $\log t$, for several different ranges, probing smaller and smaller values of $t$. We also show with the dashed line, the one-dimensional behaviour obtained from the quantum graph one-dimensional limit in (\ref{weyl1}). These plots show clearly that at very small $t$, with a finite number of iterations of the fractal, the heat kernel trace indicates truly one-dimensional behaviour. As the number of iterations increases, we see a decrease in the value of $t$ at which this one-dimensional behaviour sets in. This is to be expected. But the surprising thing is that there is a well-defined range of $t$ in which the finitely-iterated ``approximate fractal'' behaves as if it has spectral dimension $d_s$. Not only does $d_s/2$ [rather than $1/2$] give the correct overall scaling of the heat kernel trace, we also see the log periodic oscillations, indicative of ``would-be'' complex poles in the zeta function, even though the approximate fractal is a one-dimensional graph. The scale at which this occurs is determined by the size of the smallest length scale of the fractal at that level of iteration. Furthermore, this is clearly visible, for some window of $t$, with only 5 fractal iterations. This suggests that a physical probe of these fractal properties might be possible  by choosing a propagating probe of the appropriate wavelength to match this scale, in a region in which the approximate fractal looks like an ideal fractal. Similar results have been obtained \cite{dalnegro,levi}, both numerically and experimentally, for Fibonacci stacks of dielectrics \cite{fibonacci}, which are spatially regular but spectrally fractal. 
This general question of when an approximate fractal behaves like an ideal fractal is a crucial physical question, and it  deserves further mathematical attention, both analytic and numerical, in order to be made more precise. 

\section{Off-diagonal Heat Kernel Propagator}

Another vitally important problem for which there are only partial mathematical results for fractals is the question of how the well-known heat kernel propagator in (\ref{free-propagator}, \ref{liyau}) generalizes to fractals. Here the main rigorous result \cite{barlow-perkins,carpet,barlow}  gives upper and lower bounds indicating sub-diffusive propagation:
\begin{eqnarray}
\hskip -2cm \frac{c_1}{t^{d_s/2}}\, \exp\left[-c_2 \frac{(\rho(x, y))^{\frac{d_w}{d_w-1}}}{t^{\frac{1}{d_w-1}}}\right] \leq P_t(x, y)\leq  \frac{c_3}{t^{d_s/2}}\, \exp\left[-c_4 \frac{(\rho(x, y))^{\frac{d_w}{d_w-1}}}{t^{\frac{1}{d_w-1}}}\right] 
\label{bb}
\end{eqnarray}
for some real positive constants $c_1$, \dots, $c_4$, with $d_w\geq 2$ being the walk dimension and $\rho(x, y)$ an appropriate geodesic distance between $x$ and $y$. Very little is known about how  $P_t(x, y)$ behaves between these bounds, although it can be shown to be a log periodic function of the ratio $\rho(x, y)^{\frac{d_w}{d_w-1}}/t^{\frac{1}{d_w-1}}$ \cite{benarous,grigoryan}. Taking the functional form of one of these bounds as an averaged  asymptotic form for $P_t(x, y)$, we learn something interesting about the form of the Green's function at short distances. Recalling the relation (\ref{green1}) between $G(x, y)$ and $P_t(x, y)=K(x, y; t)$, we deduce the asymptotic form
\begin{eqnarray}
G(x, y)\asymp \frac{c}{(\rho(x, y))^{d_w(d_s-2)/2}}=\frac{c}{(\rho(x, y))^{d_h-d_w}}\qquad , \qquad d_s\neq 2
\label{green-fractal}
\end{eqnarray}
Thus, the familiar Green's function exponent $(d-2)$ in (\ref{green2}) for a smooth $d$-dimensional manifold, is replaced on a fractal by 
$(d_h-d_w)$, or equivalently by $d_w(d_s-2)/2$. This shows that for propagation of diffusion on a fractal the critical dimension is when the spectral dimension $d_s$ [not the Hausdorff dimension $d_h$] is equal to $2$. Implications for statistical mechanics and Bose-Einstein condensation have been studied recently by Chen \cite{chen}. In this critical case, we would expect logarithmic behaviour instead of (\ref{green-fractal}).

\section{Conclusion}

In conclusion, the zeta function and  heat kernel continue to be useful spectral functions in describing physical particles or fields propagating on fractal spaces. However, they exhibit some interesting new features on fractals, quite different from their behaviour on smooth manifolds, and these features lead to new mathematics and potentially interesting new physics.

\bigskip

Acknowledgements: I gratefully acknowledge support from the US DOE  grant DE-FG02-92ER40716, and collaboration with Eric Akkermans and Sasha Teplyaev.

\end{document}